\documentclass[conference]{IEEEtran}
\IEEEoverridecommandlockouts
\usepackage[utf8]{inputenc}
\usepackage[T1]{fontenc}
\usepackage{graphicx}
\usepackage{dblfloatfix}
\usepackage{filecontents}
\usepackage[noadjust]{cite}
\usepackage[hidelinks]{hyperref}
\usepackage{balance}
\usepackage{flushend}
\usepackage{amsmath,amssymb,amsfonts}
\usepackage{url}
\usepackage{algorithmic}
\usepackage{textcomp}
\usepackage{xcolor}
\usepackage{mdframed}
\usepackage{enumitem}
\usepackage[american]{babel}
\usepackage{subcaption}
\usepackage{multirow}
\usepackage{csquotes}
\usepackage{listings}
\usepackage[skins]{tcolorbox}

\usepackage{fancyhdr}
\fancypagestyle{copyright}{\fancyhf{}\fancyfoot[L]{This is the author’s version of the work. It is posted here for your personal use. The definitive version was published in Proceedings of ACM/IEEE International Symposium on Empirical Software Engineering and Measurement (ESEM 2023).}}
\newtcbox\tcbtp{hbox, on line, colback=lightgray, enhanced, frame hidden, boxrule=-1pt, 
    top=-2pt, bottom=-2pt, right=-1pt, left=-1pt}

\definecolor{link}{rgb}{0.63, 0.79, 0.95}

\def\BibTeX{{\rm B\kern-.05em{\sc i\kern-.025em b}\kern-.08em
    T\kern-.1667em\lower.7ex\hbox{E}\kern-.125emX}}

\begin{document}

\title{Divide and Conquer the EmpiRE:\\ A Community-Maintainable Knowledge Graph of Empirical Research in Requirements Engineering
}

\author{\IEEEauthorblockN{Oliver Karras\IEEEauthorrefmark{1}, Felix Wernlein\IEEEauthorrefmark{2}, Jil Klünder\IEEEauthorrefmark{2} and Sören Auer\IEEEauthorrefmark{1}\IEEEauthorrefmark{2}}
\IEEEauthorblockA{\IEEEauthorrefmark{1}TIB - Leibniz Information Centre for Science and Technology, Hannover, Germany\\
Email: \{oliver.karras, soeren.auer\}@tib.eu}
\IEEEauthorblockA{\IEEEauthorrefmark{2}Leibniz University Hannover, Hannover, Germany\\
Email: felix.wernlein@stud.uni-hannover.de, jil.kluender@inf.uni-hannover.de, auer@l3s.de}}

%\metatitle*{Divide and Conquer the EmpiRE: A Community-Maintainable Knowledge Graph of Empirical Research in Requirements Engineering}
%\metaauthor*{\uri{https://orcid.org/0000-0001-5336-6899}{Oliver Karras}}
%\metaauthor*{Felix Wernlein}
%\metaauthor*{\uri{https://orcid.org/0000-0001-7674-2930}{Jil Klünder}}
%\metaauthor*{\uri{https://orcid.org/0000-0002-0698-2864}{Sören Auer}}
%\researchfield*{\uri{https://orkg.org/resource/R140}{Software Engineering}}

% \author{\IEEEauthorblockN{Anonymous author 1\IEEEauthorrefmark{1}, Anonymous author 2\IEEEauthorrefmark{2}, Anonymous author 3\IEEEauthorrefmark{2} and Anonymous author 4\IEEEauthorrefmark{1}\IEEEauthorrefmark{2}}
% \IEEEauthorblockA{\IEEEauthorrefmark{1}Anonymous affiliation 1\\
% Anonymous emails}
% \IEEEauthorblockA{\IEEEauthorrefmark{2}Anonymous affiliation 2\\
% Anonymous emails}}

\definecolor{link}{RGB}{0, 123, 255}

%\IEEEpubid{978-1-5386-2880-5/17/\$31.00~\copyright~2017 IEEE}

%\IEEEoverridecommandlockouts
%\IEEEpubid{\makebox[\columnwidth]{978-1-6654-5223-6/23/\$31.00~\copyright2023 IEEE \hfill}
%\hspace{\columnsep}\makebox[\columnwidth]{ }}

\maketitle

%\IEEEpubidadjcol

%\researchproblem*{unavailability of the extracted and analyzed data from literature reviews}
%\objective*{using the Open Research Knowledge Graph (ORKG) to build and publish a community-maintainable knowledge graph whose data is openly available in the long term.}
%\method*{evaluation against competency questions}

\begin{abstract}
[Background.] Empirical research in requirements engineering (RE) is a constantly evolving topic, with a growing number of publications. Several papers address this topic using literature reviews to provide a snapshot of its ``current'' state and evolution. However, these papers have never built on or updated earlier ones, resulting in overlap and redundancy. The underlying problem is the unavailability of data from earlier works. Researchers need technical infrastructures to conduct sustainable literature reviews. [Aims.] We examine the use of the Open Research Knowledge Graph (ORKG) as such an infrastructure to build and publish an initial Knowledge Graph of Empirical research in RE (KG-EmpiRE) whose data is openly available. Our long-term goal is to continuously maintain KG-EmpiRE with the research community to synthesize a comprehensive, up-to-date, and long-term available overview of the state and evolution of empirical research in RE. [Method.] We conduct a literature review using the ORKG 
%\method{literature review using the ORKG}
to build and publish KG-EmpiRE which we evaluate against competency questions derived from a published vision of empirical research in software (requirements) engineering for 2020 -- 2025. [Results.] 
%\result{From 570 papers of the IEEE International Requirements Engineering Conference (2000 -- 2022), we extract and analyze data on the reported empirical research} 
From 570 papers of the IEEE International Requirements Engineering Conference (2000 -- 2022), we extract and analyze data on the reported empirical research and 
%\result{answer 16 out of 77 competency questions}
answer 16 out of 77 competency questions. These answers show a positive development towards the vision, but also the need for future improvements. [Conclusions.] 
%\conclusion{The ORKG is a ready-to-use and advanced infrastructure to organize data from literature reviews as knowledge graphs}
The ORKG is a ready-to-use and advanced infrastructure to organize data from literature reviews as knowledge graphs. The resulting knowledge graphs make the data openly available and maintainable by research communities, enabling sustainable literature reviews.
\end{abstract}


\section{Introduction}
\thispagestyle{copyright}
Empirical research in requirements engineering (RE) is a constantly evolving topic, with a growing number of publications~\cite{Condori.2012, Daneva.2014, Ambreen.2018}.
Several publications examined how empirical research in RE is conducted and how it should~be conducted in the future~\cite{Daneva.2014, Napoleao.2022}. Over the years, they presented snapshots of the ``current'' state and evolution of empirical research in RE~\cite{Condori.2012, Daneva.2014, Ambreen.2018, Goeken.2010, Paech.2005} and, more generally, in software engineering (SE)~\cite{Glass.2002, Jeffery.2002, Segal.2005, Zannier.2006, Hoefer.2007, Borges.2015, Zhang.2018, Molleri.2019, Guevara-Vega.2021}. They share the same goal of synthesizing a comprehensive, up-to-date, and long-term available overview of the state and evolution of empirical research in RE and SE. Although they share the same goal, use similar methods, i.a., (systematic) literature reviews, and even examine overlapping periods, venues, and themes (cf. \tablename{~\ref{tab:rw}})~\cite{Karras.2023}, they have not collaborated to build on and update earlier works, which are known challenges of literature reviews~\cite{Oelen.2021a, Oelen.2021b, Bano.2014, dos_Santos.2021}. Overcoming these challenges is critical to ensure the quality, reliability, and timeliness of research results from literature reviews~\cite{Bano.2014, Mendes.2020}.

Recent research addresses these challenges by focusing on when and how to update (systematic) literature reviews in SE and its subfields~\cite{Mendes.2020, Wohlin.2020, Napoleao.2022, Felizardo.2020}. While these works mainly provide social and economic decision support and guidance for updating literature reviews~\cite{Napoleao.2022, dos_Santos.2021}, the underlying problem is the unavailability of the extracted and analyzed data, corresponding to open science in SE~\cite{Felizardo.2020, Mendez.2020}. Unavailable data complicates collaboration among researchers and updating literature reviews, as the entire data collection, extraction, and analysis must be repeated and expanded for comprehensive results. Researchers need support in the form of technical infrastructures and services to conduct sustainable literature reviews so that all data is openly available in the long term~\cite{Goeken.2010, dos_Santos.2021, Oelen.2021a, Oelen.2021b} according to the Findable, Accessible, Interoperable, and Reusable (FAIR) data principles~\cite{Wilkinson.2016, Stocker.2023}. For this purpose, the data must be organized in a flexible, fine-grained, context-sensitive, and semantic representation to be understandable, processable, and usable by humans and machines~\cite{Auer.2020, Goeken.2010, Zhang.2018}. Over the last decade, Knowledge Graphs (KGs) have become an emerging technology in industry and academia as they enable this versatile data representation~\cite{Auer.2018, Dessi.2022, Hussein.2022}. Besides well-known KGs for encyclopedic and factual data, such as \href{https://www.dbpedia.org/}{\textit{\textcolor{link}{DBpedia}}}~\cite{Auer.2007} and \href{https://www.wikidata.org}{\textit{\textcolor{link}{WikiData}}}~\cite{Vrandevcic.2014}, using so-called Research Knowledge Graphs (RKGs) for scientific data is a rather new approach~\cite{Auer.2018, Ammar.2018, Dessi.2022}. RKGs include bibliographic metadata, e.g., titles, authors, and venues, as well as scientific data, e.g., research designs, methods, and results~\cite{Jaradeh.2019, Jeschke.2020, Paperswithcode.2022, Penev.2019, Gkatzelis.2021, Spadaro.2022}. They are a promising technology to sustainably organize scientific data so that the data is openly available for long-term collaborations~\cite{Auer.2020, Stocker.2022}.

We examine the use of RKGs as technical infrastructure by building, publishing, and evaluating an initial KG of Empirical research in RE \mbox{(KG-EmpiRE)}. Similar to Frattini et al.~\cite{Frattini.2022}, our long-term goal is to continuously maintain, (re-)use, update, and expand KG-EmpiRE with the research community to synthesize a comprehensive, up-to-date, and long-term~available overview of the state and evolution of empirical research~in RE. In this way, we can \textit{divide} the efforts to \textit{conquer} the \mbox{EmpiRE}. We use the \href{https://orkg.org/}{\textit{\textcolor{link}{Open~Research~Knowledge~Graph}}} (ORKG), a RKG with services that combine manual crowdsourcing and automated approaches to organize scientific data~\cite{Auer.2020}. Karras et al.~\cite{Karras.2021} have successfully used the ORKG to organize qualitative and quantitative \mbox{data~\cite{Karras.2021a, Karras2021b}} from two systematic literature reviews in~CrowdRE~\cite{Santos.2019, Khan.2019}.

Based on 570 papers from the \textit{IEEE International Requirements Engineering Conference} (2000 -- 2022), we show how scientific data on empirical research in RE can be consistently organized in the ORKG. In this way, we build and publish the initial KG-EmpiRE that the research community can constantly maintain, (re-)use, update, and expand. Similar to Abualhaija et al.~\cite{Abualhaija.2022}, we evaluate \mbox{KG-EmpiRE} by analyzing its data to provide initial insights into the state and evolution of empirical research in RE. In particular, we answer competency questions~\cite{Gruninger.1995, Hogan.2021} derived from the vision by Sj{\o}berg et al.~\cite{Sjoberg.2007} of how researchers should conduct empirical research in all fields of SE, including RE, in the period 2020 -- 2025. The initial insights show a positive development towards the vision~\cite{Sjoberg.2007}, but also the need for future improvements. We provide the following contributions:
%insights into using RKGs as technical infrastructure to support sustainable literature reviews with all available data and, thus, ensure the quality, reliability, and timeliness of their research results for the long-term collaboration of researchers.

\begin{mdframed}
    \begin{itemize}[leftmargin=-2.5mm]
        \textbf{Contribution:}
    	\item[] 1) The openly available KG-EmpiRE which the research community can maintain, (re-)use, update, and expand. 
        \item[] 2) A reusable and expandable ORKG \textit{template} for organizing scientific data on empirical research.
        \item[] 3) A set of 77 competency questions, the answers to which provide insights into the state and evolution of empirical research in SE and its subfields for the period 2020 -- 2025.
        \item[] 4) A reproducible data analysis of KG-EmpiRE to answer 16 of 77 competency questions, providing initial insights into the state and evolution of empirical research in RE.
    \end{itemize}
\end{mdframed}

This paper is structured as follows: Section~\ref{sec:background} explains the background. Section~\ref{sec:related_work} discusses related work. While Section~\ref{sec:approach} describes the approach, Section~\ref{sec:results} reports the results. Section~\ref{sec:threats} discusses threats to validity, and we discuss the findings in Section~\ref{sec:discussion}. Section~\ref{sec:conclusion} concludes the paper.

\section{Background}\label{sec:background}
A Research Knowledge Graph (RKG) represents scientific data semantically, i.e., explicitly and formally, by linking (meta-)data of scientific artifacts (publications, datasets, and software) and entities (persons and organizations), which offers several benefits~\cite{Auer.2018}. According to Auer et al.~\cite{Auer.2018}, the semantic representation leads to \textit{better identification}, \textit{traceability}, and \textit{reduced ambiguity} of concepts and relationships of scientific data through terminological and conceptual clarity across disciplines. These improvements result in \textit{easier \mbox{(re-)use}} of scientific data and thus \textit{less redundancy and duplication}. For example, extracted and analyzed data of literature reviews can be continuously (re-)used, updated, and expanded over time~\cite{Stocker.2023}. In addition, access to scientific data is easier for humans and especially for machines, as machines can grasp and understand the structure and semantics of scientific data in publications, so-called \textit{machine actionability}. This improved access enables far-reaching opportunities for the \textit{development of novel digital services} in science, such as customizable visualizations~\cite{Wiens.2020} and question answering systems~\cite{Jaradeh.2020, Auer.2023}.

There are \textit{generic} and \textit{specific} RKGs~\cite{Stocker.2022, Dessi.2022}. While generic RKGs focus on bibliographic metadata, e.g., titles, authors, and venues, specific RGKs focus on scientific data, e.g., research designs, methods, measurements, and results.

Generic RKGs focus on bibliographic metadata of scientific artifacts and entities. There are several well-known generic RKGs, such as \href{https://makg.org/}{\textit{\textcolor{link}{Microsoft Academic Knowledge Graph}}}~\cite{Faerber.2019}, \href{https://openalex.org/}{\textit{\textcolor{link}{OpenAlex}}}~\cite{Priem.2022}, \href{https://github.com/springernature/scigraph}{\textit{\textcolor{link}{Springer Nature SciGraph}}}~\cite{Hammond.2017}, \href{https://semanticscholar.org}{\textit{\textcolor{link}{Semantic Scholar Literature Graph}}}~\cite{Ammar.2018}, \href{https://graph.openaire.eu/}{\textit{\textcolor{link}{OpenAIRE ResearchGraph}}}~\cite{Manghi.2019, Schirrwagen.2013}, \href{https://researchgraph.org/}{\textit{\textcolor{link}{Research Graph}}}~\cite{Aryani.2017}, and \href{https://scholia.toolforge.org/}{\textit{\textcolor{link}{Scholarly Link Exchange}}} (Scholix)~\cite{Burton.2017}. These RKGs have in common that they use bibliographic metadata to organize scientific artifacts, entities, and their relationships to enable, for example, their search, visualization, and processing~\cite{Stocker.2022, Brack.2022}.

Specific RKGs focus mainly on scientific data combined with bibliographic metadata to describe and link scientific artifacts and entities. Specific RKGs are either specific to certain topics or more general to certain domains. Some well-known examples of topic-specific RKGs are \href{https://covidgraph.org/}{\textit{\textcolor{link}{CovidGraph}}}~\cite{Domingo-Fernandez.2020}, \href{https://covid-aqs.fz-juelich.de/}{\textit{\textcolor{link}{COVID-19 Air Quality Data Collection}}}~\cite{Covid-19_air_quality.2021, Gkatzelis.2021}, and \href{https://data.gesis.org/softwarekg/}{\textit{\textcolor{link}{SoftwareKG}}}~\cite{Schindler.2020, Schindler.2021, Schindler.2022}. The first two RKGs address the topic of COVID-19. However, the \textit{CovidGraph} looks more generally at scientific data on COVID-19 to explore publications, patents, existing treatments, and drugs around the coronavirus family~\cite{Domingo-Fernandez.2020}. In contrast, the \textit{COVID-19 Air Quality Data Collection} is more fine-grained, focusing only on scientific content from publications about the impacts of COVID-19 lockdowns on air quality~\cite{Covid-19_air_quality.2021, Gkatzelis.2021}. The \textit{SoftwareKG} deals with the topic of software that is mentioned in scientific publications. This RKG enables users to explore and understand the role of software in science. Besides topic-specific RKGs, there are several examples of domain-specific RKGs~\cite{Stocker.2022}, such as \href{https://scholkg.kmi.open.ac.uk/}{\textit{\textcolor{link}{Computer Science Knowledge Graph}}} (CS-KG)~\cite{Dessi.2022}, \href{https://paperswithcode.com/}{\textit{\textcolor{link}{Papers-with-Code}}}~\cite{Paperswithcode.2022}, \href{https://hi-knowledge.org/}{\textit{\textcolor{link}{Hi Knowledge}}}~\cite{Jeschke.2020}, \href{https://cooperationdatabank.org/}{\textit{\textcolor{link}{Cooperation Databank}}} (CoDa)~\cite{Spadaro.2022}, and \href{http://openbiodiv.net/}{\textit{\textcolor{link}{OpenBiodiv}}}~\cite{Penev.2019}. These RGKs organize scientific data from a specific domain, including computer science~\cite{Dessi.2022}, machine learning~\cite{Paperswithcode.2022}, invasion biology~\cite{Jeschke.2020}, social sciences~\cite{Spadaro.2022}, and biodiversity~\cite{Penev.2019}.

\begin{figure*}[!t]
    \captionsetup{justification=justified}
    \centering
    \includegraphics[width=0.75\textwidth]{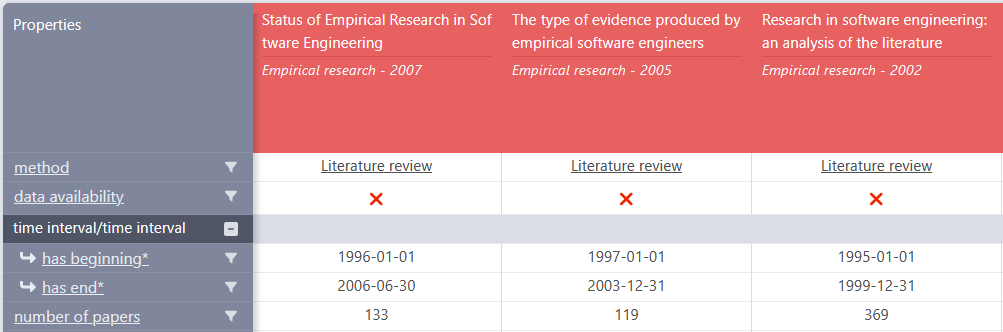}
    \caption{\href{https://doi.org/10.48366/r255464}{\textit{\textcolor{link}{Comparison}}} of related publications on the ``current'' state and evolution of empirical research in RE and SE~\cite{Karras.2023}.} \label{comparison}
    \vspace{-0.5cm}
\end{figure*}

\begin{table*}[!b]
	\captionsetup{justification=centering, singlelinecheck=off}
	\caption{Details of related publications on the ``current'' state and evolution of empirical research in RE and SE~\cite{Karras.2023}.\\\footnotesize{Legend: Literature Review (LR), Systematic Literature Review (SLR), and Systematic Mapping Study (SMS)}}
	\centering
	\label{tab:rw}
	\resizebox{\textwidth}{!}{%
		\begin{tabular}{|c|c|c|l|l|r|l|l|l|}
			\hline
			\textbf{Paper} & \textbf{Year} & \textbf{Field} & \multicolumn{1}{c|}{\textbf{Method}} & \multicolumn{1}{c|}{\textbf{Period}} & \multicolumn{1}{c|}{\textbf{Data basis}} & \multicolumn{1}{c|}{\textbf{Dataset}} & \multicolumn{1}{c|}{\textbf{Venues} (Frequency $> 2$ )} & \multicolumn{1}{c|}{\textbf{Themes} (Frequency $> 2$)}\\ \hline \hline
			\cite{Paech.2005} & 2005 & \multirow{6}{*}{RE} & LR & 1986 -- 2002 & 35 papers & Unavailable & \multirow{15}{*}{\parbox{6.5cm}{\begin{enumerate}[leftmargin=3.75mm]
						\item Empirical Software Engineering Journal (8)
						\item IEEE Software (4)
						\item Requirements Engineering Journal (4)
						\item ACM/IEEE International Symposium on Empirical Software Engineering and Measurement (4)
						\item IEEE Transactions on Software Engineering (3)
						\item Information and Software Technology Jounral (3)
						\item IEEE International Requirements Engineering Conference (3)
						\item Journal of Systems and Software (3)
						\item International Conference on Software Engineering (3)
			\end{enumerate}}} & \multirow{15}{*}{\parbox{3.6cm}{\begin{enumerate}[leftmargin=3.75mm]
						\item Data collection (12)
						\item Research method (11)
						\item Bibliographic metadata (10)
						\item Data analysis (8)
						\item Research paradigm (7)
						\item Research design (7)
						\item Research topic (5)
						\item Research context (4)
						\item Sample of population (4)
						\item Theory (3)
			\end{enumerate}}} \\ \cline{1-2} \cline{4-7}
			\cite{Goeken.2010} & 2010 &  & SLR & Unknown & 154 papers & Broken link &  &  \\ \cline{1-2} \cline{4-7}
			\cite{Condori.2012} & 2012 &  & Survey & \begin{tabular}[c]{@{}l@{}}19.03.2012 -- \\ 30.03.2012\end{tabular} & 42 respondents & Unavailable &  &  \\ \cline{1-2} \cline{4-7}
			\cite{Daneva.2014} & 2014 &  & LR & 1983 -- 2013 & 2237 papers & Unavailable &  &  \\ \cline{1-2} \cline{4-7}
			\cite{Ambreen.2018} & 2016 &  & SMS & Open -- 2012 & 270 papers & Unavailable &  &  \\ \cline{1-7} 
			\cite{Glass.2002} & 2002 & \multirow{9}{*}{SE} & LR & 1995 -- 1999 & 369 papers & Unavailable &  &  \\ \cline{1-2} \cline{4-7}
			\cite{Jeffery.2002} & 2002 & & LR & 1996 -- 2002 & 68 papers & Unavailable &  &  \\ \cline{1-2} \cline{4-7}
			\cite{Segal.2005} & 2005 &  & LR & 1997 -- 2003 & 119 papers & Unavailable &  &  \\ \cline{1-2} \cline{4-7}
			\cite{Zannier.2006} & 2006 &  & LR & 1977 -- 2005 & 63 papers & Unavailable &  &  \\ \cline{1-2} \cline{4-7}
			\cite{Hoefer.2007} & 2007 &  & LR & 1996 -- 2003 & 133 papers & Unavailable &  &  \\ \cline{1-2} \cline{4-7}
			\cite{Borges.2015} & 2015 &  & SMS & 1996 -- 2013 & 891 papers & Broken link &  &  \\ \cline{1-2} \cline{4-7}
			\cite{Zhang.2018} & 2018 &  & SMS & 2013 -- 2017 & 538 papers & Broken link &  &  \\ \cline{1-2} \cline{4-7}
			\cite{Molleri.2019} & 2019 &  & SMS & 1991 -- 2014 & 341 papers & Unavailable &  &  \\ \cline{1-2} \cline{4-7}
			\cite{Guevara-Vega.2021} & 2021 &  & SMS & Open -- 2019 & 20 papers & Available &  &  \\ \hline
		\end{tabular}
	}
\end{table*}

In contrast to all RKGs mentioned, the ORKG is a special case as it is a RKG organizing any topic-specific scientific data across all research domains. Thus, the ORKG is a cross-domain and cross-topic RKG. The ORKG organizes scientific data provided by a publication as a collection of so-called \textit{contributions}. A contribution consists of a semantic description of scientific data. Selected contributions can be compared in so-called \textit{comparisons}. A comparison is a table where the columns denote the selected contributions by publication and the rows denote the semantically described scientific~data.
\figurename{~\ref{comparison}} shows an excerpt from a comparison that we created to get an overview of related publications on the ``current'' state and evolution of empirical research in RE and SE~\cite{Karras.2023}. For three publications, the excerpt shows the method used, the data availability, as well as the period and the number of papers examined. We use the ORKG due to its cross-domain and cross-topic characteristics, as well as its successful application for CrowdRE by Karras et al.~\cite{Karras.2021}.

\section{Related Work}\label{sec:related_work}
Below, we review 14 publications that provide snapshots of the ``current'' state and evolution of empirical research in RE and SE (see \tablename{~\ref{tab:rw}})~\cite{Karras.2023}. We only consider publications that address the topic in general and are not limited to specific aspects, such as a method~\cite{Sjoberg.2005, Bezerra.2015} or a context~\cite{Dyba.2008, Zhang.2016}.

We found five publications on empirical research in RE published between 2005 and 2016 and nine on empirical research in SE published between 2002 and 2021. While one publication~\cite{Condori.2012} examined empirical research in RE using a survey with 42 respondents, the other 13 publications~\cite{Paech.2005, Goeken.2010, Daneva.2014, Ambreen.2018, Glass.2002, Jeffery.2002, Segal.2005, Zannier.2006, Hoefer.2007, Borges.2015, Zhang.2018, Molleri.2019, Guevara-Vega.2021} used (systematic) literature reviews or systematic mapping studies to analyze on average 402.9 papers (minimum: 20, median: 154, and maximum: 2237 papers) published between 1977 and 2019 with overlapping periods. In total, these 13 publications examined papers from a total of 60 different venues on 18 different themes. Nine of the 60 venues and ten of the 18 themes were examined by more than two publications. These facts show that there is considerable overlap and redundancy between these publications in terms of goals, methods used, periods, venues, and themes examined. This overlap and redundancy could have been avoided if researchers had collaborated to build on and update earlier works. However, only four out of 14 publications offer their data at all, with only one publication~\cite{Guevara-Vega.2021} using a public data repository~\cite{Guevara.2021a, Guevara.2021b}. The other three publications only offer links that no longer work~\cite{Goeken.2010, Borges.2015, Zhang.2018}.

In terms of key findings, the 14 publications show consistent results, although not all 18 themes were examined in all publications. For example, eleven of the 14 publications reported on the most commonly used research methods. Until 2000, the most common research methods were conceptual analysis and concept implementation~\cite{Glass.2002}. Between 2000 and 2015, the most commonly used research methods changed to case studies and experiments~\cite{Jeffery.2002, Ambreen.2018, Goeken.2010, Hoefer.2007, Segal.2005, Zannier.2006}, which were expanded after 2015 to also include surveys and systematic literature reviews~\cite{Guevara-Vega.2021, Molleri.2019, Borges.2015, Zhang.2018}. While this change shows an evolution of research methods used, we also note that experiments and case studies have been the two main research methods for empirical research in RE and SE for more than 20 years. Although these two research methods have been used for a long time, seven publications concluded that there is a need to develop, expand, and use standardized terminology and theories (from other disciplines) to more consistently represent the empirical research conducted and better explain the results found~\cite{Daneva.2014, Condori.2012, Paech.2005, Zhang.2018, Borges.2015, Glass.2002, Jeffery.2002}. In this regard, seven publications also analyzed the information reported for a comprehensive description of a research design. This information includes details about the research question(s)~\cite{Condori.2012}, contextual factors~\cite{Paech.2005}, object of study~\cite{Condori.2012}, population/sample~\cite{Ambreen.2018, Zannier.2006}, threats to validity~\cite{Zhang.2018}, data collection~\cite{Guevara-Vega.2021}, measurements/metrics~\cite{Hoefer.2007}, and data analysis~\cite{Zhang.2018, Zannier.2006}. Overall, six publications~\cite{Ambreen.2018, Daneva.2014, Paech.2005, Zhang.2018, Borges.2015, Zannier.2006} explicitly state that the use of empirical research in RE and SE is constantly increasing. This increase is accompanied by the need to have a comprehensive, up-to-date, and long-term available overview of the state and evolution of empirical research~\cite{Condori.2012, Goeken.2010, Paech.2005}. This need is the underlying motivation of all 14 related publications.

We build on the related work by considering their examined venues and themes for our data collection, extraction, and analysis. Unlike related work, we do not conduct a full systematic literature review or systematic mapping study. Instead, we conduct a literature review to illustrate how researchers can use RKGs, specifically the ORKG, as a technical infrastructure for organizing scientific data in an openly available and long-term way to build and publish KGs that the research community can constantly maintain, (re-)use, update, and expand. We do not claim to provide a comprehensive overview of the state and evolution of empirical research in RE. Our research approach aims to lay the foundation for such an overview by building, publishing, and evaluating the initial KG-EmpiRE.

\section{Research Approach}\label{sec:approach}
The research approach and reporting essentially follow the \textit{Empirical Standards for Software Engineering Research}~\cite{Ralph.2021}.

We first define the research goal and research question to ensure that the scope of our research approach is clearly defined before presenting its details. We defined the research goal in detail using the goal definition template~\cite{Basili.1994}:

\begin{mdframed}
    \begin{itemize}[leftmargin=-2.5mm]
    	\item[] \textbf{Goal definition:} We \textit{analyze} the ORKG \textit{for the purpose of} organizing scientific data in an openly available and long-term way \textit{with respect to} building, publishing, and evaluating an initial KG of empirical research in RE that the research community can constantly maintain, (re-)use, update, and expand, \textit{from the point of view of} ORKG users \textit{in the context of} enabling sustainable literature reviews to synthesize a comprehensive, up-to-date, and long-term available~overview of the state and evolution of empirical research~in RE.
    \end{itemize}
\end{mdframed}

Based on this goal, we ask the following research question:
\begin{mdframed}
    \begin{itemize}[leftmargin=-2.5mm]
    	\item[] \textbf{Research question:} How can we use the ORKG as a technical infrastructure to organize scientific data in an openly available and long-term way by building, publishing, and evaluating a KG of empirical research in RE that the research community can constantly maintain, \mbox{(re-)use}, update, and expand to enable sustainable literature reviews?
    \end{itemize}
\end{mdframed}

\begin{figure}[htbp]
    \captionsetup{justification=justified}
    \centering
    \includegraphics[width=\columnwidth]{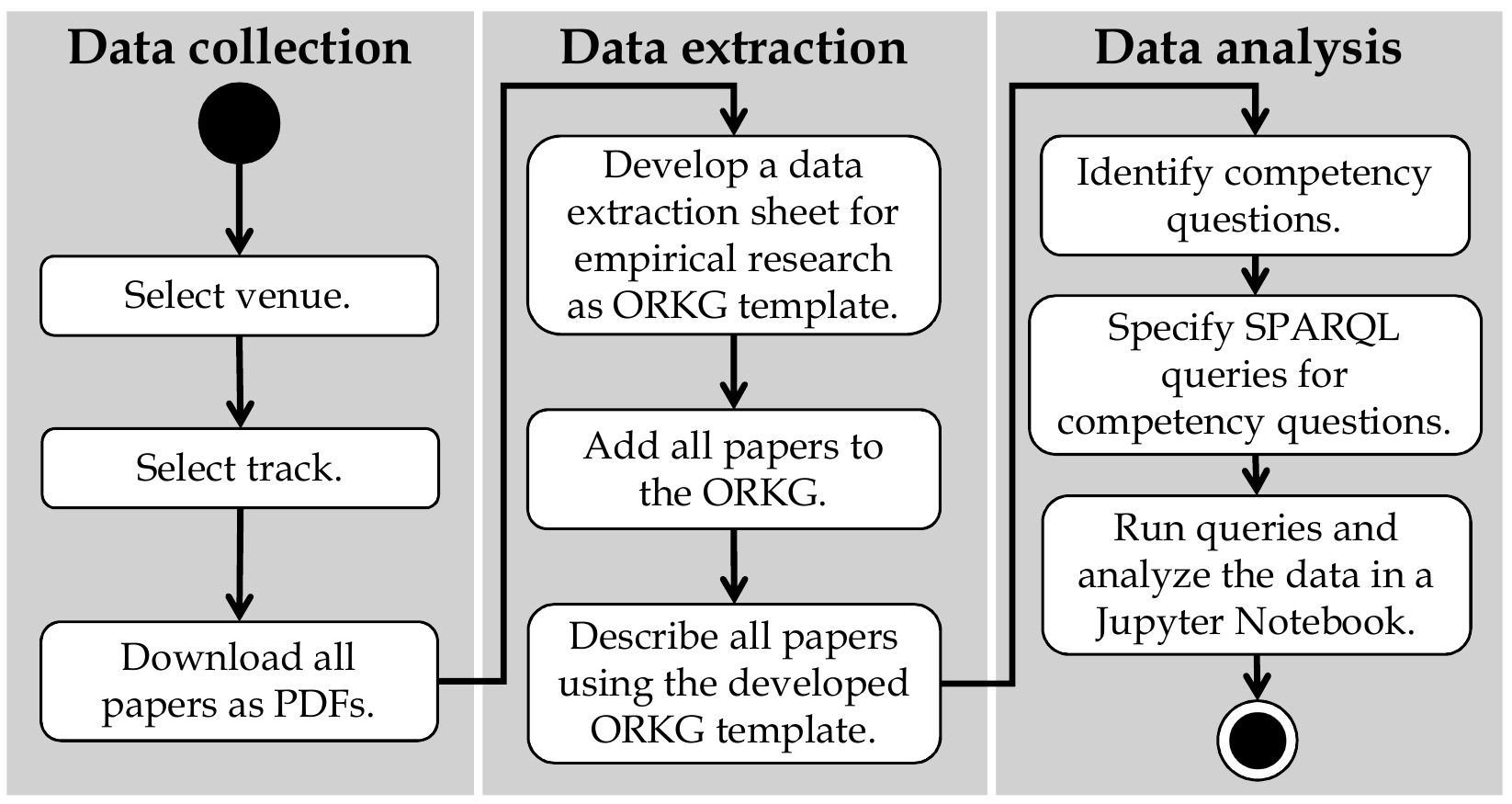}
    \caption{Activity diagram of the research approach.} \label{approach}
    \vspace{-0.5cm}
\end{figure}

We frame our research approach using the design science paradigm~\cite{Runeson.2020}. We \textit{collect} papers from the field of RE and \textit{extract} data on the empirical research reported. In this way, we build and publish the initial KG-EmpiRE as a solution design, which we \textit{analyze} for evaluation. For this reason, our research approach consists of the three main steps: Data collection, data extraction, and data analysis (see \figurename{~\ref{approach}}).

\subsection{Data Collection}
\label{sec:data_collection}
Many related publications collected data on empirical research in RE and SE from papers (cf. Section~\ref{sec:related_work}). However, the only available data set is small and covers only a subset of the frequently examined themes (cf.~\tablename{~\ref{tab:rw}})~\cite{Karras.2023}. For this reason, we conducted another data collection (see \figurename{~\ref{approach}}).

We considered only papers from one venue similar to other related publications~\cite{Hoefer.2007, Segal.2005, Guevara-Vega.2021, Zannier.2006, Jeffery.2002, Goeken.2010} to simplify the search and selection. The selected venue is the \textit{IEEE International Requirements Engineering Conference}, as authors from related publications reported that most of the papers they identified as relevant came from this conference~\cite{Daneva.2014, Ambreen.2018}. We selected the research track of the conference, as it is the main track where we expect most papers applying empirical research\footnote{The selected venue and track are only a starting point for our work. We know that the other tracks of the \textit{IEEE International Requirements Engineering Conference} are also relevant and that there are other important venues on the topic of empirical research in RE, such as the \href{https://link.springer.com/journal/10664/volumes-and-issues}{\textit{\textcolor{link}{Empirical Software Engineering}}} journal, \href{https://www.computer.org/csdl/magazine/so/past-issues/2020/2023}{\textit{\textcolor{link}{IEEE Software}}} journal, or \href{https://link.springer.com/journal/766/volumes-and-issues}{\textit{\textcolor{link}{Requirements Engineering}}} journal, etc. (cf.~\tablename{~\ref{tab:rw}})~\cite{Karras.2023}.}.

\begin{table*}[!b]
\centering
\caption{Overview of a subset of the content identified for data extraction.}
\label{tab:content}
\resizebox{.85\textwidth}{!}{%
    \begin{tabular}{|p{2.5cm}||l|l|l|}
        \hline
        \multicolumn{1}{|c||}{\textbf{Related Publications}} & \multicolumn{1}{c|}{\textbf{Theme}} & \multicolumn{1}{c|}{\textbf{Analyzed content}} & \multicolumn{1}{c|}{\textbf{Possible values}} \\ \hline \hline
        \multirow{3}{*}{\parbox{2.5cm}{\cite{Ambreen.2018, Daneva.2014, Paech.2005, Zhang.2018, Borges.2015, Hoefer.2007, Segal.2005, Glass.2002, Molleri.2019, Guevara-Vega.2021, Zannier.2006, Jeffery.2002}}} & \multirow{3}{*}{Data collection} & \multirow{2}{*}{Data} & Type of data: Qualitative, Quantitative \\ \cline{4-4}
        %&  &  & Source of data: Open-Source, Industry, Academia, etc. \\ \cline{4-4}
        &  &  & Location of the data: URL \\ \cline{3-4} 
        &  & \multirow{2}{*}{Collection method} & \multirow{2}{*}{Case study, Survey, Interview, Experiment, etc.}  \\ \cline{1-2}
        \multirow{2}{*}{\parbox{2.5cm}{\cite{Ambreen.2018, Goeken.2010, Zhang.2018, Borges.2015, Hoefer.2007, Segal.2005, Glass.2002, Molleri.2019, Guevara-Vega.2021, Zannier.2006, Jeffery.2002}}} & \multirow{2}{*}{Research method} &  &  \\ \cline{3-4} 
        &  & \multirow{2}{*}{Analysis method} & \multirow{2}{*}{Comparative analysis, Content analysis, Grounded theory, etc.} \\ \cline{1-2}
        \multirow{3}{*}{\parbox{2.5cm}{\cite{Condori.2012, Paech.2005, Zhang.2018, Hoefer.2007, Segal.2005, Glass.2002, Molleri.2019, Zannier.2006}}} & \multirow{3}{*}{Data analysis} &  & \\ \cline{3-4} 
        &  & Inferential statistic & Statistical test: t-test, ANOVA, Logistic regression, etc. \\ \cline{3-4} 
        &  & Descriptive statistic & Type of measure: Frequency, Central tendency, Position, etc. \\ \hline
        % & \multirow{2}{*}{Machine learning} & Algorithm: Naive Bayes, SVM, Logistic Regression, etc. & \multirow{2}{*}{} \\ \cline{3-3}
        % &  & Metric: Precision, Recall, F1-score, Accuracy, etc. &  \\ \hline
        \multirow{2}{*}{\parbox{2.5cm}{\cite{Ambreen.2018, Paech.2005, Zhang.2018, Segal.2005, Glass.2002, Guevara-Vega.2021, Jeffery.2002}}} & \multirow{2}{*}{Research paradigm} & \multirow{2}{*}{Type of paradigm} & \multirow{2}{*}{Exploratory, Explanatory, Descriptive, Predictive, etc.} \\
        & & & \\
        \hline
        \multirow{4}{*}{\parbox{2.5cm}{\cite{Ambreen.2018, Condori.2012, Paech.2005, Zhang.2018, Molleri.2019, Guevara-Vega.2021, Zannier.2006}}} & \multirow{4}{*}{Research design} & \multirow{3}{*}{Research question} & Formulated question \\ \cline{4-4}
        &  &  & Presentation: Explicitly highlighted, Implicitly hidden  \\ \cline{4-4}
        &  &  & Type of question: Exploratory, Explanatory, Descriptive, etc.  \\ \cline{3-4} 
        %&  & \multirow{2}{*}{Hypothesis} & Formulated hypothesis \\ \cline{4-4}
        %&  &  & Type of hypothesis: Null, Alternative \\ \cline{3-4}
        &  & Threats to Validity & Type of validity: Internal, External, Construct, Conclusion, etc. \\ \hline 
        % & Research answer & Visibility: Explicitly highlighted, Implicitly hidden & \cite{Condori.2012} \\ \hline
        \parbox{2.5cm}{\cite{Ambreen.2018, Daneva.2014, Zhang.2018, Borges.2015, Hoefer.2007, Segal.2005, Glass.2002, Molleri.2019, Zannier.2006, Jeffery.2002}} & Bib. metadata & Metadata & Title, Authors, Venue, Publication date, URL, DOI \\ \hline
    \end{tabular}%
    }
\end{table*}

For data collection, we downloaded the papers as PDF files from \href{https://ieeexplore.ieee.org/xpl/conhome/1000630/all-proceedings}{\textit{\textcolor{link}{IEEE Xplore}}}, where all proceedings of the \textit{IEEE International Requirements Engineering Conference} from 1994 -- 2022 can be found. So far, we downloaded all 570 papers from the research track of the conference from 2000 -- 2022 for data extraction. The collection of the missing years (1994 -- 1999) is future work. We did not need to exclude any paper, as each paper reported at least partial information about empirical research that we could extract.

\subsection{Data Extraction}
The data extraction is the essential step in building and publishing the initial KG-EmpiRE. The basis of data extraction are the related publications with their data extraction sheets, themes, and contents (cf. \tablename{~\ref{tab:rw}})~\cite{Karras.2023}.

We focused on the themes examined in more than five related publications. These themes are \textit{data collection}, \textit{research method}, \textit{bibliographic metadata}, \textit{data analysis}, \textit{research paradigm}, and \textit{research design}. For each theme, we analyzed the related publications to determine the analyzed content and possible values (see \tablename{~\ref{tab:content}}). In this way, we determined the content for the data extraction covering the most frequently examined themes. \tablename{~\ref{tab:content}} provides an overview of a subset of the content identified for data extraction. For a complete overview of all content identified for data extraction, refer to our \href{https://github.com/okarras/EmpiRE-Analysis/blob/master/Supplementary\%20materials/Overview\%20of\%20all\%20content\%20for\%20data\%20extraction.pdf}{\textit{\textcolor{link}{supplementary materials}}}~\cite{Karras.2023a}.

Instead of a spreadsheet, we implemented the data extraction sheet as a so-called ORKG \textit{template}~\cite{Kabongo.2021} to organize the scientific data (see \figurename{~\ref{approach}}). ORKG templates are an implementation of a subset of SHACL~\cite{Knublauch.2017} and allow specifying the structure of ORKG \textit{contributions} to describe a paper (cf.~Section~\ref{sec:background}). In this way, we determined which data to extract and ensured that all the semantic descriptions of scientific data are consistent and comparable across all considered papers.

\begin{figure}[!ht]
    \captionsetup{justification=justified}
    \centering
    \includegraphics[width=1.0\columnwidth]{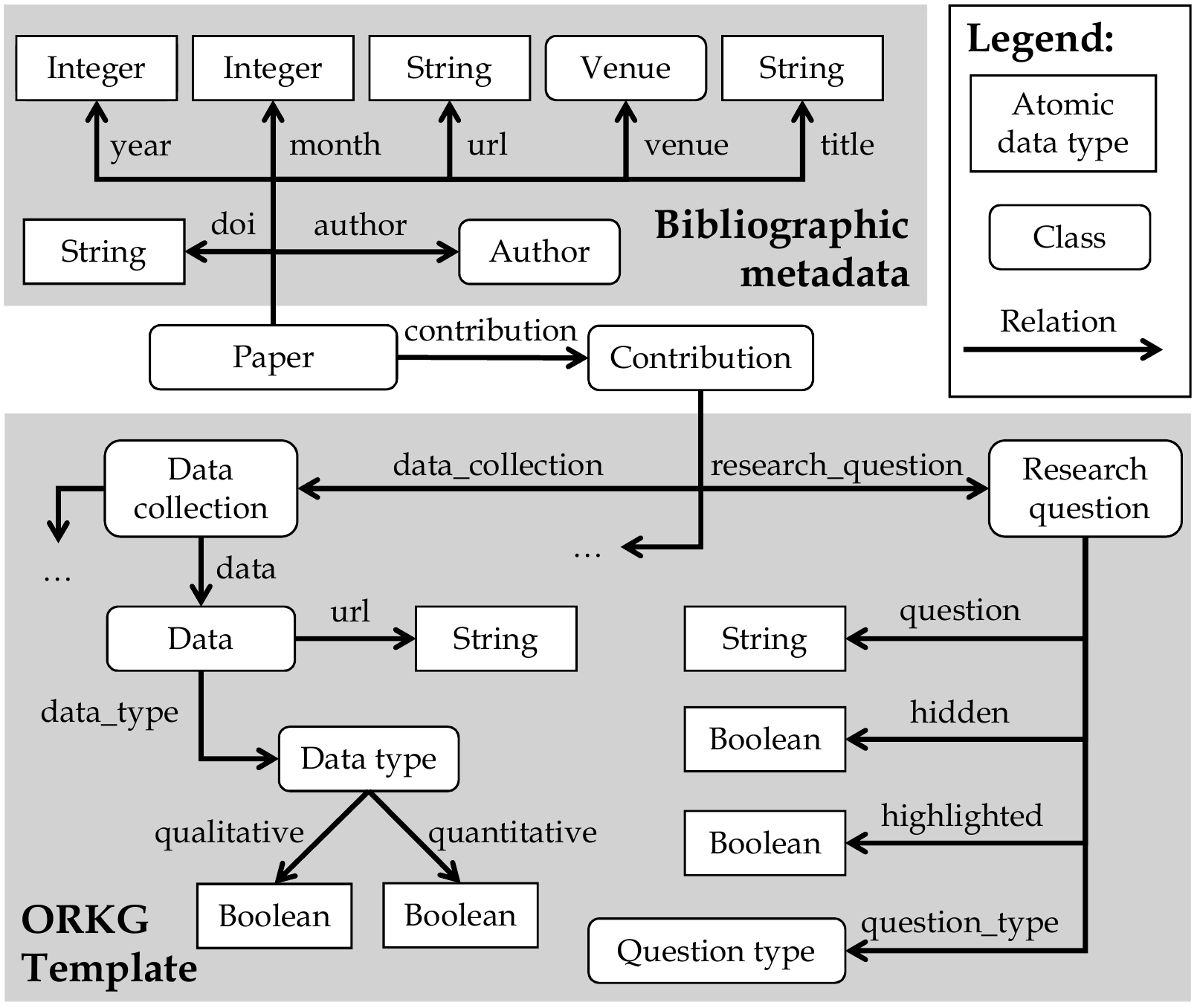}
    \caption{Excerpt from the ORKG \textit{template} for data extraction.} \label{template}
    \vspace{-0.5cm}
\end{figure}

The first two authors of this paper developed the 
\href{https://orkg.org/template/R186491}{\textit{\textcolor{link}{ORKG template}}} in four iterations over a period of two months. When developing the ORKG \textit{template}, we focused on a generic design to ensure its reusability. Starting from an initial draft, we applied the (revised) template to five randomly selected papers from the data collection in each iteration. Based on our experiences in data extraction, we continuously adapted the template and always updated the descriptions of all papers from which we had previously extracted data. After the fourth iteration, there were no more changes. The remaining two authors reviewed the final version of the template and confirmed its suitability for data extraction. \figurename{~\ref{template}} shows an excerpt from the ORKG \textit{template} to illustrate the structure for describing the data collected (theme: \textit{data collection}) and research question(s) posed (theme: \textit{research design}) in a paper. Our ORKG \textit{template} excludes bibliographic metadata, as the ORKG \mbox{(semi-)automatically} compiles the bibliographic metadata of a paper when the paper is added. For an overview of the ORKG \textit{template}, refer to our \href{https://github.com/okarras/EmpiRE-Analysis/blob/master/Supplementary\%20materials/Detailed\%20ORKG\%20template\%20structure.pdf}{\textit{\textcolor{link}{supplementary materials}}}~\cite{Karras.2023a}.

\begin{table*}[!b]
\centering
\caption{Three examples of derived competency questions (cf. \href{https://github.com/okarras/EmpiRE-Analysis/blob/master/Supplementary\%20materials/Detailed\%20list\%20of\%20all\%2077\%20competency\%20questions.xlsx}{\textit{\textcolor{link}{supplementary materials}}}~\cite{Karras.2023a}).}
\label{tab:competency-questions}
\resizebox{\textwidth}{!}{%
   \begin{tabular}{|c|l|ll|c|}
        \hline
        \multirow{2}{*}{\textbf{ID}} & \multicolumn{1}{c|}{\multirow{2}{*}{\textbf{Competency question}}} & \multicolumn{2}{c|}{\textbf{Excerpts from the vision of Sj{\o}berg et al.}~\cite{Sjoberg.2007}} & \multicolumn{1}{c|}{\multirow{2}{*}{\begin{tabular}[c]{@{}c@{}}\textbf{Related publications}\\\textbf{with a similar question}\end{tabular}}} \\ \cline{3-4} 
         
        & \multicolumn{1}{|c|}{} & \multicolumn{1}{c|}{\textbf{State of practice (2007)}} & \multicolumn{1}{c|}{\textbf{Target state (2020 -- 2025)}} & \\ \hline \hline
         
        \multirow{3}{*}{1}& \multicolumn{1}{|l|}{\multirow{3}{*}{\parbox{4.6cm}{How has the proportion of empirical studies evolved over time?}}} & \multicolumn{1}{l|}{\multirow{3}{*}{\parbox{5cm}{There are relatively \tcbtp{\textit{few empirical studies}.}}}} & \multicolumn{1}{p{5cm}|}{A \tcbtp{\textit{large number of studies}} covering all important fields of SE and using different empirical methods are conducted and reported.} & \multirow{3}{*}{\parbox{2.5cm}{\centering\cite{Daneva.2014, Ambreen.2018, Zannier.2006}}} \\ \hline
        
        \multirow{3}{*}{26} & \multicolumn{1}{|p{4.6cm}|}{How have the proportions of case studies and action research in the empirical methods used evolved over time?} & \multicolumn{1}{l|}{\multirow{3}{*}{\parbox{5cm}{One may question the \tcbtp{\textit{industrial relevance}} \tcbtp{\textit{of most SE studies}}.}}} & \multicolumn{1}{l|}{\multirow{3}{*}{\parbox{5cm}{\tcbtp{\textit{More case studies and action research}} should be carried out.}}} & \multirow{3}{*}{\parbox{2.5cm}{\centering\cite{Ambreen.2018, Jeffery.2002, Zannier.2006, Borges.2015, Zhang.2018}}} \\ \hline
        
        \multirow{3}{*}{39} & \multicolumn{1}{|p{4.6cm}|}{How has the provision of data (the materials used, raw data collected, and study results identified) evolved over time?} & \multicolumn{1}{l|}{\multirow{3}{*}{\parbox{5cm}{\tcbtp{\textit{Few studies provide results that enable}} \tcbtp{\textit{efficient cumulative research}}, [\dots]}}} & \multicolumn{1}{l|}{\multirow{3}{*}{\parbox{5cm}{\tcbtp{\textit{More research studies}} are designed~with the goal of \tcbtp{\textit{enabling efficient use of its}} \tcbtp{\textit{results}} by other researchers.}}} & \multirow{3}{*}{\parbox{2.5cm}{\centering None.}} \\ \hline
    \end{tabular}%
}
\end{table*}

For data extraction, we added each paper from data collection to the ORKG using its Digital Object Identifier (DOI). In this way, the ORKG automatically compiles the bibliographic metadata of the papers. The second author applied the developed ORKG \textit{template} to each paper and extracted the corresponding data from all papers, using the terminology used in the paper to ensure an accurate and consistent description. The first and the third author reviewed each description by comparing the extracted data with the respective paper. In the case of inconsistencies or ambiguities, the three authors discussed and resolved the issues identified.

\subsection{Data Analysis}
The data analysis serves two purposes: \hypertarget{P1}{(P1) We evaluate the coverage of the curated topic of empirical research in RE by the initial KG-EmpiRE}, and \hypertarget{P2}{(P2) We get initial insights into the state and evolution of empirical research in RE}.

Competency questions are an established method for analyzing and evaluating KGs~\cite{Gruninger.1995}. A competency question is a natural language question that represents an information need related to the content of a KG and for which a KG must provide relevant information to answer the question~\cite{Hogan.2021}. Thus, the number of questions answered reflects the coverage of the curated topic in a KG \hyperlink{P1}{(P1)}, and the answers to competency questions provide insights into the curated topic \hyperlink{P2}{(P2)}.

Following our approach (see \figurename{~\ref{approach}}), we identified relevant competency questions about the state and evolution of empirical research. We selected the vision of Sj{\o}berg et al.~\cite{Sjoberg.2007} regarding the role of empirical methods in all fields of SE, including RE, to identify competency questions, as the vision precisely targets the current period of 2020 -- 2025. Sj{\o}berg et al.~\cite{Sjoberg.2007} present their vision by describing and comparing the ``current'' state of practice (2007) and their targeted state (2020 -- 2025). %for five specific themes: Extent of empirical studies, quality of empirical studies, relevance of empirical studies, synthesis of evidence, and the use of theory.
The first three authors analyzed these descriptions by manually coding them in terms of textual elements that led to a question related to the state and evolution of empirical research. In this way, we derived the competency questions and captured their associated origins so that third parties better understand our analysis and its results. Subsequently, the first three authors matched each identified competency question with the analyzed questions and content of the related publications~(cf. Section~\ref{sec:related_work}) to determine how many times similar questions have been analyzed. \tablename{~\ref{tab:competency-questions}} shows three examples of identified competency questions, including excerpts from the vision~\cite{Sjoberg.2007} with light gray highlighting of the coded text elements that led to the question and the references to related publications with a similar question. In total, we identified 77 competency questions about the state and evolution of empirical research in SE, including RE, of which 42 questions have been asked similarly in at least one related publication. For the detailed list of all 77 competency questions, refer to our \href{https://github.com/okarras/EmpiRE-Analysis/blob/master/Supplementary\%20materials/Detailed\%20list\%20of\%20all\%2077\%20competency\%20questions.xlsx}{\textit{\textcolor{link}{supplementary materials}}}~\cite{Karras.2023a}.

For data analysis, we specified queries with a query language for KGs to retrieve the extracted data from the ORKG. This specification requires knowledge of the data structure, i.e., the ORKG \textit{template}, so we can only specify queries for competency questions that can be answered with the extracted data. We used the query language SPARQL~\cite{Seaborne.2013} as the ORKG provides a SPARQL endpoint for accessing all its data. Listing~\ref{lst:sparql} exemplary shows the specified SPARQL query for competency question 1 (cf. \tablename{~\ref{tab:competency-questions}}). In this case, we present the query with human-readable identifiers to facilitate understanding the query. This human-readable query is not executable in the ORKG as the ORKG uses alphanumeric identifiers (similar to \textit{WikiData}~\cite{Vrandevcic.2014}). For the executable queries, refer to our \href{https://github.com/okarras/EmpiRE-Analysis/blob/master/empire-analysis.ipynb}{\textit{\textcolor{link}{supplementary materials}}}~\cite{Karras.2023a}.

\begin{footnotesize}
\begin{lstlisting}[language=SPARQL, caption=SPARQL query for competency question 1., xleftmargin=0.5cm, label={lst:sparql}]
PREFIX r: <http://orkg.org/orkg/resource/>
PREFIX c: <http://orkg.org/orkg/class/>
PREFIX p: <http://orkg.org/orkg/predicate/>
PREFIX rdfs: <http://www.w3.org/2000/01/rdf-schema\#>

SELECT ?paper, ?year, ?dc_label, ?da_label
WHERE {
    ?paper p:contribution ?contri;
           p:publication_year ?year.
    ?contri. a c:C27001.
    OPTIONAL{?contri p:data_collection_method ?dc.
             ?dc rdfs:label ?dc_label.}
    OPTIONAL{?contri p:data_analysis_method ?da.
             ?da rdfs:label ?da_label.}
}
\end{lstlisting}
\end{footnotesize}

We implemented the analysis using a \href{https://github.com/okarras/EmpiRE-Analysis/blob/master/empire-analysis.ipynb}{\textit{\textcolor{link}{Jupyter Notebook}}}~\cite{Karras.2023a} with \textit{Python}, published on \href{https://github.com/okarras/EmpiRE-Analysis}{\textit{\textcolor{link}{GitHub}}}~\cite{Karras.2023a} with the latest release archived on \href{https://zenodo.org/record/8083529}{\textit{\textcolor{link}{Zenodo}}}~\cite{Karras.2023a}. We also hosted the repository on \href{https://mybinder.org/v2/gh/okarras/EmpiRE-Analysis/HEAD?labpath=\%2Fempire-analysis.ipynb}{\textit{\textcolor{link}{mybinder}}}~\cite{Karras.2023a}. In this way, the analysis is always available to anyone for interactive reproduction and (re-)use, retrieving the latest data from the ORKG. Due to the uniform data structure provided by the developed ORKG \textit{template}, we can always retrieve newly added papers that use the ORKG \textit{template} and include them in the analysis by simply rerunning the script.

\section{Results}\label{sec:results}
Below, we present the results of the data analysis. First, we address the coverage of the curated topic by the initial \mbox{KG-EmpiRE} \hyperlink{P1}{(P1)}, followed by initial insights into the state and evolution of empirical research in RE based on the competency questions answered \hyperlink{P2}{(P2)}.

\subsection{Coverage of the Curated Topic by KG-EmpiRE}
Overall, we answered 16 of the 77 competency questions (21\%) using KG-EmpiRE. 13 of these 16 competency questions were asked similarly in at least one related publication.

Given the initial stage of KG-EmpiRE, the number of competency questions answered represents an acceptable coverage of the curated topic. So far, we have focused only on one track of one venue, and the ORKG \textit{template} covers only six of the 18 different themes examined in related publications (cf. Section~\ref{sec:related_work}). Therefore, the need to expand KG-EmpiRE by adding more papers from different venues and more data is evident. In particular, the organization of additional data to answer the open competency questions is necessary, which requires expanding the ORKG \textit{template}. However, our goal was not to build and publish an already extensive KG of empirical research in RE to answer as many competency questions as possible. Instead, we aimed to lay its foundation by building, publishing, and evaluating the initial KG-EmpiRE. We conducted a literature review to illustrate~how researchers can use RKGs, specifically the ORKG, as a technical infrastructure for organizing scientific data in an openly available and long-term way to build and publish KGs that the research community can constantly maintain, (re-)use, update, and expand.

\subsection{State and Evolution of Empirical Research in RE}
Due to the initial stage of KG-EmpiRE, we cannot provide an analysis of the general state and evolution of empirical research in RE. Nevertheless, we show some results from the data analysis so far. Due to space limitations, we present only specific results for the three competency questions from \tablename{~\ref{tab:competency-questions}} before reporting more generally on our initial insights.\vspace{0.1cm}

\noindent
\textit{CQ1: How has the proportion of empirical studies evolved?}

According to the \href{https://www.springer.com/journal/10664}{\textit{\textcolor{link}{Empirical Software Engineering}}} journal, ``\textit{Empirical studies presented here usually involve the collection and analysis of data and experience} [\dots]". For this reason, we define that an empirical study always includes data collection and analysis. For each year, we examine the relative proportion of all papers that meet the definition to all papers collected, as the absolute number of papers varies per year.

\begin{figure*}[!htbp]
    \captionsetup{justification=justified}
    \centering
    \includegraphics[width=0.8\textwidth]{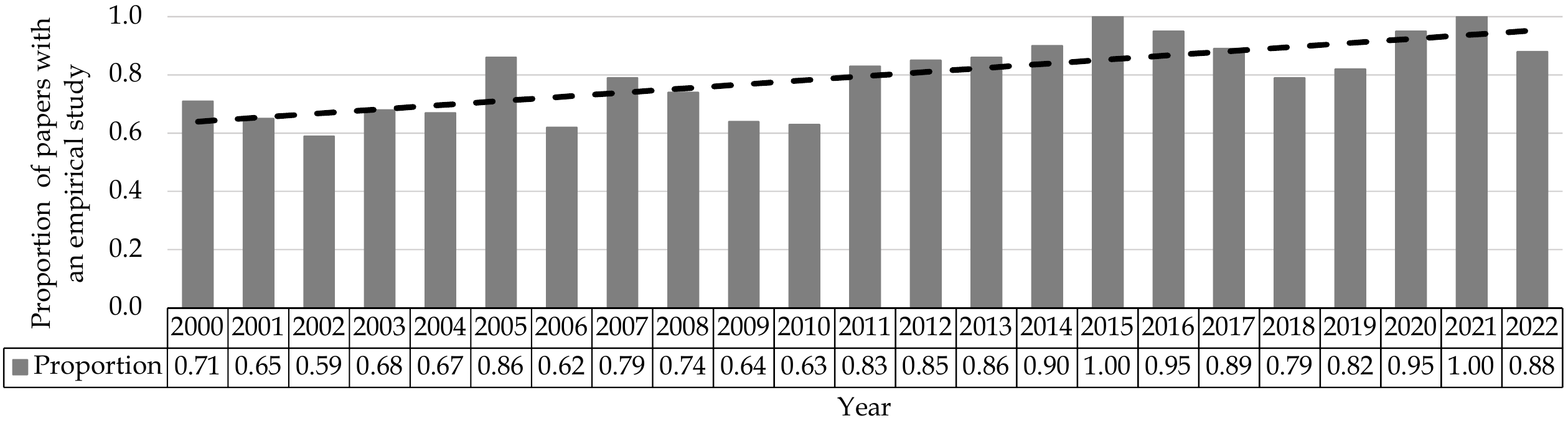}
    \caption{Proportion of papers per year reporting an empirical study with a linear trend line.}\label{CQ1}
    \vspace{-0.51cm}
\end{figure*}

\figurename{~\ref{CQ1}} shows that the proportion of papers that report an empirical study is always above 58\%, averaging 79.6\%, and increases slightly over time. While before 2010 the average proportion is 69.5\%, the average proportion for the period 2010 -- 2019 is 85.2\%. For the target state of the vision (2020 -- 2025)~\cite{Sjoberg.2007}, the average proportion is 94.3\%. Based on these results, we conclude a positive development towards the vision so that a large number of empirical studies can be achieved.\vspace{0.1cm}

\noindent
\textit{CQ26: How have the proportions of case studies and action research in the empirical methods used evolved?}

For each year, we examine the relative proportion of all papers that report \textit{case study} or \textit{action research} as a data collection method to all papers reporting a data collection method, as the absolute number of papers varies per year.

\begin{figure*}[htbp]
    \captionsetup{justification=justified}
    \centering
    \includegraphics[width=0.8\textwidth]{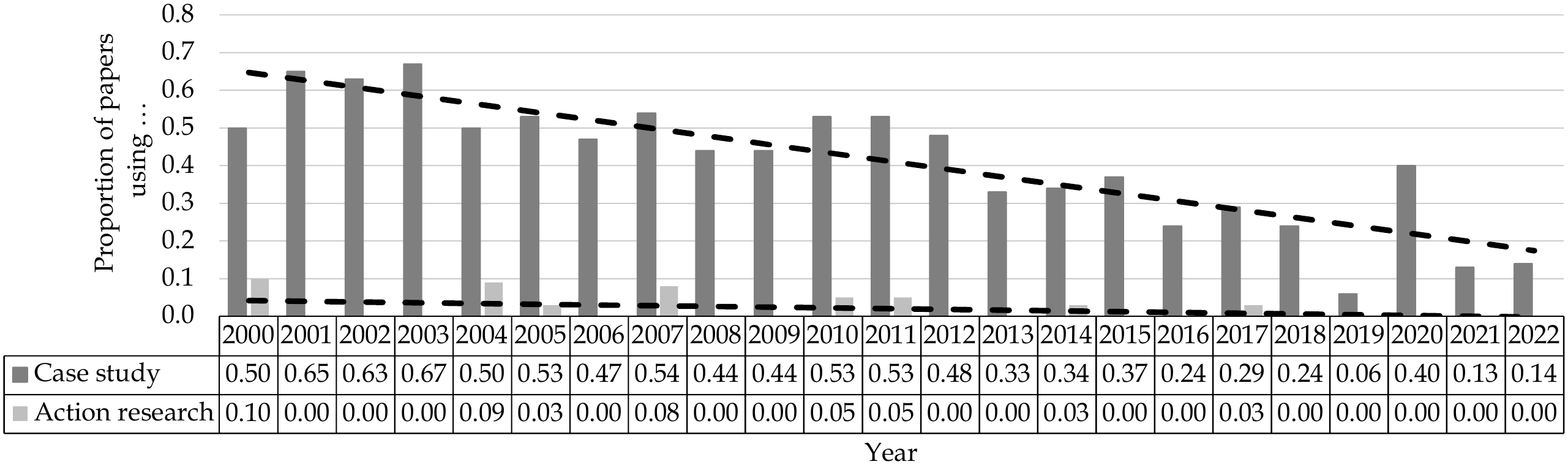}
    \caption{Proportion of papers per year using case study or action research with linear trend lines.}\label{CQ26}
    \vspace{-0.51cm}
\end{figure*}

In \figurename{~\ref{CQ26}}, we present the proportion of papers per year that report \textit{case study} or \textit{action research}, each of which we address individually below. The proportion of papers using \textit{case study} decreases over time, averaging 41.1\%. While before 2010 the average proportion of papers is 53.7\%, the average proportion for the period 2010 -- 2019 is 34.1\%. For the target state of the vision (2020 -- 2025)~\cite{Sjoberg.2007}, the average proportion is 23.6\%. We assume this decrease is due to a better understanding of the term ``\textit{case study}'' among researchers. A recent study by Wohlin~\cite{Wohlin.2021} found that researchers often misused the term ``\textit{case study}'' in software engineering research. We can confirm this finding as several papers analyzed use the term ``\textit{case study}'', although, at best, they report an experiment or a larger use case. Despite the decrease, this finding represents a positive development of the empirical research in RE, as researchers make better use of the term ``\textit{case study}''.
The proportion of papers using \textit{action research} is constantly low over time. In only eight years, papers use \textit{action research} at all and the proportions are at most 10\%, averaging 2\%. Since 2018, no paper uses \textit{action research}, so the average proportion for the target state of the vision (2020 -- 2025)~\cite{Sjoberg.2007} is 0\%.
Based on these results, we conclude that the increased use of case studies and action research, as envisioned for the target state, has not yet been achieved.\vspace{0.1cm}

\noindent
\textit{CQ39: How has the provision of data (the materials used, raw data collected, and study results identified) evolved?}

For each year, we examine the relative proportion of all papers that report at least one URL to their \textit{data} to all papers reporting a data collection method, as the absolute number of papers varies per year.

\begin{figure*}[htbp]
    \captionsetup{justification=justified}
    \centering
    \includegraphics[width=0.8\textwidth]{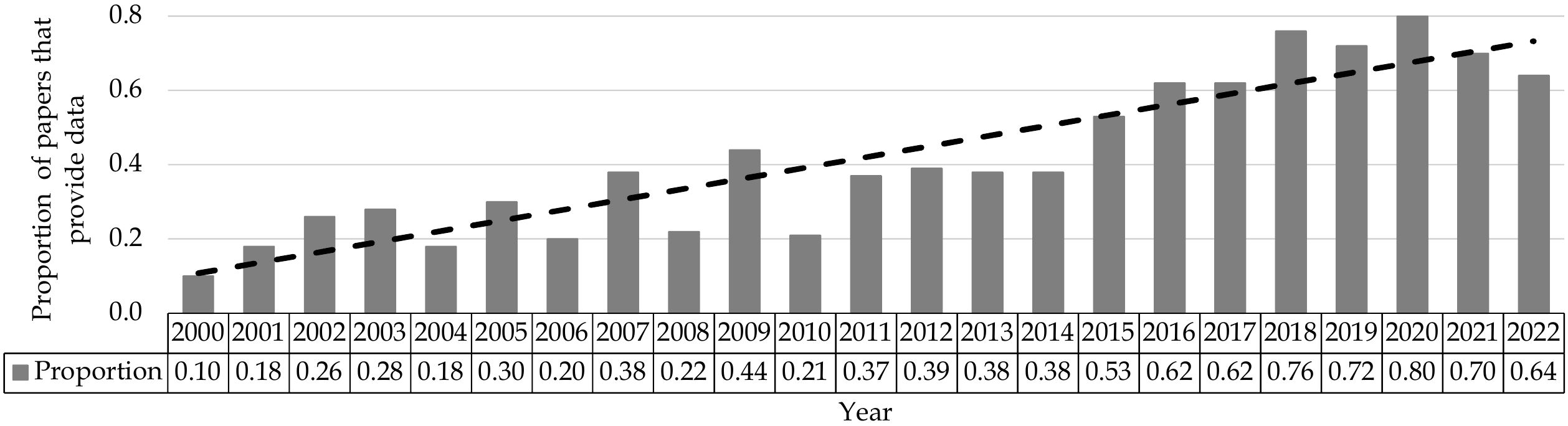}
    \caption{Proportion of papers per year that report at least one URL to their data with a linear trend line.}\label{CQ39}
    \vspace{-0.51cm}
\end{figure*}

\figurename{~\ref{CQ39}} shows that the proportion of papers that report at least one URL to their \textit{data} increases remarkably over time, averaging 42\%. While before 2010 the average proportion of papers is 25.4\%, the average proportion for the period 2010 -- 2019 is 49.8\%. For the target state of the vision (2020 -- 2025)~\cite{Sjoberg.2007}, the average proportion is 71.3\%. Based on these results, we conclude a positive development towards the vision that more empirical studies will provide their \textit{data}.\vspace{0.1cm}

\noindent
\textit{Initial insights from the data analysis}

Overall, the data analysis shows a positive development of the state and evolution of empirical research in RE towards the vision of Sj{\o}berg et al.~\cite{Sjoberg.2007}. For the following statements, we provide all associated analyses with visualizations and explanations as an interactive \href{https://mybinder.org/v2/gh/okarras/EmpiRE-Analysis/HEAD?labpath=\%2Fempire-analysis.ipynb}{\textit{\textcolor{link}{Jupyter Notebook}}}~\cite{Karras.2023a}.

We found that the proportion of papers reporting an empirical study increases over time, with an average proportion of 94.3\% for the target state (2020 -- 2025). Regarding the use of empirical methods, we observed that the number of empirical methods used for data collection and analysis in a single paper increases over time, with three to four empirical methods being most frequently used in one paper overall. For the target state, researchers mainly use three to even five empirical methods in one paper with average proportions of 22\% for three, 25.3\% for four, and 26.7\% for five empirical methods.
For data collection, researchers frequently and constantly use the established empirical methods \textit{experiment}, \textit{secondary research}, and \textit{survey}, with average proportions of 35.7\% (\textit{experiment}), 40\% (\textit{secondary research}), and 18.7\% (\textit{survey}) for the target state. We also found that the use of \textit{case study} decreases over time, with an average proportion of 22.3\% for the target state, and that \textit{action research} is rarely used with an average proportion of 0\% for the target state. For data analysis, researchers mainly and constantly use \textit{descriptive statistics} with a proportion of 87.6\% overall and 92\% for the target state. In contrast, the use of \textit{inferential statistics} with a proportion of 19.2\% overall and 26.3\% for the target state is low. We also found a positive development~in~the reporting of research design. The proportion of papers reporting \textit{threats to validity}, providing \textit{data}, and highlighting \textit{research questions and answers} steadily increase over time with average proportions of 91.3\% (\textit{threats to validity}), 71.3\% (\textit{data}), and 23.7\% (\textit{research questions and answers}) for the target~state.%\vspace{0.1cm}

Despite the positive developments towards the vision~\cite{Sjoberg.2007}, we have also identified the need for future improvements.

According to Sj{\o}berg et al.~\cite{Sjoberg.2007}, more \textit{case studies} and \textit{action research} are needed for data collection to ensure the industrial relevance of empirical research. However, our results show that the use of \textit{case studies} decreased, and researchers rarely use \textit{action research}. More effort from the research community is required to achieve this part of the vision. For data analysis, the proportion of papers using \textit{inferential statistics} is low (26.2\% average proportion for the target state) and diverse. Based on the names, we found 57 different statistical tests in the 570 papers, often apparently using different spellings for partly the supposedly same test. For example, we found at least six different spellings for the \textit{Mann-Whitney U test}~\cite{Wilcoxon.1945, Mann.1947}. For a mature use of statistical methods as envisioned by Sj{\o}berg et al.~\cite{Sjoberg.2007}, the research community needs to be more concerned with \textit{inferential statistics} and should use references to identify the statistical tests used. Regarding the reporting of \textit{threats to validity}, we found two issues for future improvements. First, a proportion of 33.6\% of the papers reporting \textit{threats to validity} did not use any further classification of the types of validity. Although the general reporting of \textit{threats to validity} is useful, the lack of classification makes it difficult for a reader to have a clear overview of whether the \textit{threats to validity} have been comprehensively discussed. Second, a proportion of 18.2\% of the papers reporting \textit{threats to validity} addresses \textit{conclusion validity}. This proportion is low compared to the proportions for other types of validity (\textit{external validity}: 60.4\%, \textit{internal validity}: 56.1\%, and \textit{construct validity}: 47.9\%). In the future, the research community needs to communicate threats to validity comprehensively and transparently by discussing all types of validity equally and naming the types of validity~addressed. 

All of these insights are initial and subject to various threats to validity, which we discuss in more detail in the following.

\section{Threats to Validity}\label{sec:threats}
We discuss threats to \textit{study selection}, \textit{data}, and \textit{research validity} based on the guideline for managing threats to validity of secondary studies in SE by Ampatzoglou et al.~\cite{Ampatzoglou.2020}.\vspace{0.1cm}

\textit{Study selection validity} includes threats to the search process and study filtering. We ensured the \textit{adequacy of initial relevant paper identification} by selecting all papers from the research track of the \textit{IEEE International Requirements Engineering Conference}. This venue is the largest international RE conference, where established RE researchers regularly publish high-quality, peer-reviewed (empirical) research~\cite{Daneva.2014, Ambreen.2018}. The \textit{selection of one publication venue} simplified the search and selection process due to the broad topic of empirical research in RE, while leading to a representative subset of relevant papers for building the initial KG-EmpiRE. However, the validity of KG-EmpiRE is limited due to the lack of relevant papers from other important venues.
%such as the journal \href{https://link.springer.com/journal/766/volumes-and-issues}{\textit{Requirements Engineering}} or the \href{https://link.springer.com/conference/refsq}{\textit{International Working Conference on Requirement Engineering: Foundation for Software Quality}}~\cite{Davis.2009}.
Collecting more papers from other venues is future work. We have been able to \textit{access all papers} from the research track of the \textit{IEEE International Requirements Engineering Conference} as our research institutes provide us access to \href{https://ieeexplore.ieee.org/xpl/conhome/1000630/all-proceedings}{\textit{\textcolor{link}{IEEE Xplore}}}, where all proceedings of the \textit{IEEE International Requirements Engineering Conference} from 1994 -- 2022 can be found. 

\textit{Data validity} includes threats to data extraction and data analysis. We mitigated \textit{data extraction bias} and \textit{researcher bias} by having a clear strategy: 1) The second author extracted the data from the papers, 2) The first and third authors reviewed the extracted data by comparing them with the papers, and 3) The three authors directly discussed and resolved any inconsistencies or ambiguities identified. For the data extraction, the first two authors developed an ORKG \textit{template} in four iterations over two months that the remaining two authors finally reviewed and confirmed. We determined the contents of the ORKG \textit{template} based on the related publications with their data extraction sheets, themes, and analyzed content (cf. \tablename{~\ref{tab:content}}) to mitigate \textit{classification schema bias}. \textit{Publication bias} is present in the initial KG-EmpiRE, as we selected only papers from one publication venue. Nevertheless, we assume that the selected papers are representative of empirical research in RE in general and thus ensure the \textit{validity of primary studies}, as the \textit{IEEE International Requirements Engineering Conference} is known for its high-quality (empirical) research.

\textit{Research validity} includes threats to the entire research process and design. This paper has a \textit{research method bias} resulting from the use of only one method for data collection, namely a literature review. Although this method has its weaknesses, its systematic reflection is a proven means for building an initial, sound knowledge base~\cite{Glinz.2015, Karras.2020}. We also explicitly state that KG-EmpiRE is initial, and our long-term goal is to continuously maintain KG-EmpiRE with the research community to synthesize a comprehensive, up-to-date, and long-term~available overview of the state and evolution of empirical research in RE. For these reasons, we assume that our research approach is appropriate to illustrate how researchers can use RKGs, specifically the ORKG, as a technical infrastructure for organizing scientific data in an openly available and long-term way to build and publish KGs that the research community can constantly maintain, (re-)use, update, and expand. We described our systematic research approach in detail and made all \href{https://github.com/okarras/EmpiRE-Analysis}{\textit{\textcolor{link}{data and materials}}}~\cite{Karras.2023a} openly available to ensure their \textit{reliability}, \textit{repeatability}, and \textit{reproducibility}. Concerning \textit{generalizability}, our initial insights are consistent with the findings of the related publications (cf. Section~\ref{sec:related_work}). However, these initial insights are limited. Despite the reputation of the \textit{IEEE International Requirements Engineering Conference} and its representativeness of the RE community, the data analysis mainly provides insights into the state and evolution of empirical research in RE published at this conference, so they do not necessarily reflect the state and evolution of empirical research in RE as a whole.

\section{Discussion}\label{sec:discussion}
Empirical research in RE is a constantly evolving topic, with a growing number of publications. The ever-growing number of publications is a well-known problem~\cite{Jinha.2010, White.2019}, as it becomes almost impossible to keep track of the current state of research~\cite{Auer.2020}. Therefore, there is a need for a comprehensive, up-to-date, and long-term available overview of the state and evolution of empirical research in RE~\cite{Condori.2012, Goeken.2010, Paech.2005}.

Over the years, several publications have addressed this need (cf. Section~\ref{sec:related_work}), but none of these publications built on or updated earlier ones, resulting in considerable overlap and redundancy between them. The underlying problem is the unavailability of data from earlier work to maintain, \mbox{(re-)}use, update, and expand them (cf. \tablename{~\ref{tab:rw}})~\cite{Karras.2023}. Researchers need technical infrastructures and services to conduct sustainable literature reviews so that all data is openly available in the long term~\cite{Goeken.2010, dos_Santos.2021, Oelen.2021a, Oelen.2021b}. In this paper, we examine the use of RKGs, specifically the ORKG, as such a technical infrastructure for building, publishing, and evaluating an initial KG of empirical research in RE, which is openly available for long-term collaboration among researchers. For this purpose, we collected 570 papers from the \textit{IEEE International Requirements Engineering Conference} and extracted data from them on the six most frequently examined themes (see \tablename{~\ref{tab:rw}})~\cite{Karras.2023}. Based on 16 answered competency questions (out of 77), we present initial insights from the analysis of the KG-EmpiRE that are consistent with the findings of related publications.

Fundamentally, our research approach (cf. \figurename{~\ref{approach}}) is the same as any literature review and consists of manual data collection, extraction, and analysis. Even if manual collection and extraction still require a lot of effort, there are first approaches~\cite{Martin.2022, Bless.2023} that enable authors to describe new papers semantically while writing. For example, authors could combine our ORKG \textit{template} with these approaches to collect the relevant data on their empirical research themselves. The resulting semantic descriptions can then be imported into RKGs, which can simplify the data collection and extraction for literature reviews in the long term and, in the best case, even make them obsolete. Bless et al.~\cite{Bless.2023} have already presented a proof-of-concept by importing and updating semantic descriptions created with their approach into the ORKG.

Despite the similarity of our approach to any literature review, there are important differences in its implementation that leads to decisive advantages. Using the ORKG, the extracted data is not encapsulated in a file as usual, which is, at best, published on a data repository, but in an openly available knowledge graph, which, to put it simply, is nothing more than a graph-based database. Overall, the ORKG offers a ready-to-use and sustainably governed infrastructure that implements best practices, such as FAIR principles~\cite{Wilkinson.2016} and versioning, with services to support researchers in organizing FAIR scientific data~\cite{Stocker.2023}. As a result, the FAIR scientific~data is openly available in the long term and can be understood, processed, and used by humans and machines. Thus, the~research community can constantly maintain, (re-)use, update, and expand the initial KG-EmpiRE, that we have built, published, and evaluated, in a long-term and collaborative~manner. For example, in case of errors in data extraction, anyone, and in the best case the authors themselves, can update the data. It is also possible to expand KG-EmpiRE by curating more papers using or even expanding our ORKG \textit{template} to extract more data in a structured, consistent, and comparable way. In all these cases, anyone can (re-)use our \href{https://github.com/okarras/EmpiRE-Analysis/blob/master/empire-analysis.ipynb}{\textit{\textcolor{link}{Jupyter Notebook}}}~\cite{Karras.2023a} to reproduce the data analysis and its (updated) results.

Based on our results, the ORKG is a promising option as a technical infrastructure for conducting sustainable literature reviews and thus also for the continuous systematic literature review (CSLR) process envisioned by Napole{\~a}o et al.~\cite[p. 6]{Napoleao.2022}. For a full implementation of the CSLR process, however, the ORKG needs additional features, such as \textit{forward snowballing} or \textit{suggestion of potentially relevant papers}. Such future use and expansion of the ORKG is realistic and feasible as the ORKG team seeks to collaborate with others on use cases and new features for their system~\cite{Karras.2021}. As an answer to our research question, we can summarize:

\begin{mdframed}
    \begin{itemize}[leftmargin=-2.5mm]
    	\item[] \textbf{Answer to the research question:} The ORKG is an innovative technical infrastructure with services that we can use \textit{directly} for organizing scientific data in an openly available and long-term way to build, publish, and evaluate a KG of empirical research in RE that the reseacher community can maintain, (re-)use, update, and expand. Moreover, due to its conception, the ORKG fundamentally enables building and publishing KGs on any topic from any domain, thus laying the foundation for sustainable literature reviews. Despite the existing functionality, there is a need to expand the ORKG to better support researchers in collecting papers, extracting data, and analyzing them. Such an expansion of the ORKG is realistic and feasible so that the ORKG even has the potential to become a suitable platform for the CSLR process.
    \end{itemize}
\end{mdframed}

%2005: 35, 2004: 27, 2003: 25, 2002: 32, 2001: 26, 2000: 14, 1999: 19, 1998: 25, 1997: 20, 1996: 30, 1995: 20, 1994: 33, 1993: 21 = 327 weitere paper, insgesamt 738
For our future work, we have a plan with short-, \mbox{mid-,} and long-term actions. As short-term action, we expand \mbox{KG-EmpiRE} by describing more papers with our ORKG \textit{template}. Our goal over the coming months is to cover the entire research track of the \textit{IEEE International Requirements Engineering Conference} from 1994 -- 2022 to get a comprehensive overview of the state and evolution of empirical research in RE at this conference. We also establish a more general ORKG \textit{observatory} on \href{https://orkg.org/observatory/Empirical_Software_Engineering}{\textit{\textcolor{link}{Empirical Software Engineering}}} as a central access point to all curated papers. The observatory is an open group that anyone can join to contribute to the topic.
As mid-term actions, we write and publish an ORKG \textit{review} about the state and evolution of empirical research in RE, based on the complete collection of all papers from the research track of the \textit{IEEE International Requirements Engineering Conference}. An ORKG \textit{review} is a special kind of literature review article that the research community can constantly maintain when underlying content in the ORKG changes due to updates or expansions~\cite{Oelen.2021a, Oelen.2021b}. Besides the ORKG \textit{review}, we also expand KG-EmpiRE by including more papers from other important venues (cf. \tablename{~\ref{tab:rw}})~\cite{Karras.2023} to gain a more comprehensive overview of the state and evolution of empirical research in RE. As long-term action, we expand our ORKG \textit{template} to organize more extensive scientific data about empirical research in a structured, consistent, and comparable manner and thus to address the 61 still open competency questions. With this plan, we work towards maintaining, updating, and expanding the initial KG-EmpiRE together with the research community by \textit{dividing} the efforts to \textit{conquer} the EmpiRE.

\section{Conclusion}\label{sec:conclusion}
Empirical research in RE is a constantly evolving topic. Several publications address this topic, i.a., using (systematic) literature reviews. However, they only provide snapshots of the ``current'' state and evoluation but no comprehensive, up-to-date, and long-term available overview of the state and evolution of empirical research in RE. The underlying problem is the unavailability of data from earlier works to build on and update collaboratively. While recent research addresses these challenges by providing social and economic decision support and guidance, researchers need technical infrastructures and services to conduct sustainable literature reviews.

We examine the use of RKGs, specifically the ORKG, as such a technical infrastructure. We conduct a literature review using the ORKG for organizing scientific data from 570 papers in an openly available and long-term way. As a result, we build and publish the initial KG-EmpiRE that the research community can constantly maintain, (re-)use, update, and expand. We analyze KG-EmpiRE for evaluation by answering 16 out of 77 competency questions derived from a published vision of empirical research in software (requirements) engineering for the period 2020 -- 2025~\cite{Sjoberg.2007}. 
Besides consistent findings with the related publications, the analysis shows a positive development towards the vision~\cite{Sjoberg.2007}, but also the need for future improvements.

We conclude that the use of the ORKG and RKGs, in general, is a step in the right direction to allow researchers to build on and update earlier works, enabling sustainable literature reviews to ensure the quality, reliability, and timeliness of research results for successful long-term collaboration among researchers. Comprehensive, up-to-date, and long-term available overviews of the state and evolution of broad topics such as empirical research in RE are major research challenges that we as a research community can only \textit{conquer} by \textit{dividing} the efforts, true to the principle: \textit{Divide et Impera}.

%\section*{Remark for Reviewers}
%We have already made the developed content and materials openly available. For this reason, we cannot guarantee that reviewers will not discover our identities if they \textbf{actively} search for the created content and materials in the Open Research Knowledge Graph (ORKG) and on GitHub themselves. We have made all content and materials available for reviewers in an \href{https://anonymous.4open.science/r/EmpiRE-Analysis-D283/}{\textit{\textcolor{link}{\underline{Anonymous GitHub}}}} repository.

\section*{Data Availability}
The data supporting the findings of this study are openly available in the \href{https://orkg.org/observatory/Empirical_Software_Engineering}{\textit{\textcolor{link}{Open Research Knowledge Graph}}}. We provide all \href{https://github.com/okarras/EmpiRE-Analysis}{\textit{\textcolor{link}{supplementary materials}}}~\cite{Karras.2023a}, a \href{https://github.com/okarras/EmpiRE-Analysis/blob/master/Supplementary\%20materials/rdf-export-orkg-2023-06-26.nt}{\textit{\textcolor{link}{snapshot of data in the ORKG}}}~\cite{Karras.2023a}, and the interactive \href{https://mybinder.org/v2/gh/okarras/EmpiRE-Analysis/HEAD?labpath=\%2Fempire-analysis.ipynb}{\textit{\textcolor{link}{Jupyter Notebook}}}~\cite{Karras.2023a} of our analysis on \href{https://github.com/okarras/EmpiRE-Analysis}{\textit{\textcolor{link}{GitHub}}} and \href{https://zenodo.org/record/8083529}{\textit{\textcolor{link}{Zenodo}}}~\cite{Karras.2023a}.

\section*{Acknowledgment}
The authors thank the Federal Government, the Heads of Government of the Länder, as well as the Joint Science Conference (GWK), for their funding and support within the NFDI4Ing and NFDI4DataScience consortia. This work was funded by the German Research Foundation (DFG) -project numbers 442146713 and 460234259, by the European Research Council for the project ScienceGRAPH (Grant agreement ID: 819536), and by the TIB - Leibniz Information Centre for Science and Technology.

\bibliographystyle{IEEEtran}
\balance
\bibliography{IEEEabrv,references}

% Generated by IEEEtran.bst, version: 1.14 (2015/08/26)
\begin{thebibliography}{10}
\providecommand{\url}[1]{#1}
\csname url@samestyle\endcsname
\providecommand{\newblock}{\relax}
\providecommand{\bibinfo}[2]{#2}
\providecommand{\BIBentrySTDinterwordspacing}{\spaceskip=0pt\relax}
\providecommand{\BIBentryALTinterwordstretchfactor}{4}
\providecommand{\BIBentryALTinterwordspacing}{\spaceskip=\fontdimen2\font plus
\BIBentryALTinterwordstretchfactor\fontdimen3\font minus
  \fontdimen4\font\relax}
\providecommand{\BIBforeignlanguage}[2]{{%
\expandafter\ifx\csname l@#1\endcsname\relax
\typeout{** WARNING: IEEEtran.bst: No hyphenation pattern has been}%
\typeout{** loaded for the language `#1'. Using the pattern for}%
\typeout{** the default language instead.}%
\else
\language=\csname l@#1\endcsname
\fi
#2}}
\providecommand{\BIBdecl}{\relax}
\BIBdecl

\bibitem{Condori.2012}
N.~Condori-Fernandez, M.~Daneva, and R.~Wieringa, ``{A Survey on Empirical
  Requirements Engineering Research Practices},'' in \emph{18th International
  Working Conference on Requirements Engineering: Foundation for Software
  Quality}, 2012.

\bibitem{Daneva.2014}
M.~Daneva, D.~Damian, A.~Marchetto, and O.~Pastor, ``{Empirical Research
  Methodologies and Studies in Requirements Engineering: How Far Did We
  Come?}'' \emph{Journal of Systems and Software}, vol.~95, 2014.

\bibitem{Ambreen.2018}
T.~Ambreen, N.~Ikram, M.~Usman, and M.~Niazi, ``{Empirical Research in
  Requirements Engineering: Trends and Opportunities},'' \emph{Requirements
  Engineering}, vol.~23, no.~1, 2018.

\bibitem{Napoleao.2022}
B.~M. Napole{\~a}o, F.~Petrillo, S.~Hall{\'e}, and M.~Kalinowski, ``{Towards
  Continuous Systematic Literature Review in Software Engineering},'' in
  \emph{48th Euromicro Conference on Software Engineering and Advanced
  Applications}.\hskip 1em plus 0.5em minus 0.4em\relax IEEE, 2022.

\bibitem{Goeken.2010}
M.~Goeken and J.~Patas, ``{Evidence-Based Structuring and Evaluation of
  Empirical Research in Requirements Engineering},'' \emph{Business \&
  Information Systems Engineering}, vol.~2, no.~3, 2010.

\bibitem{Paech.2005}
B.~Paech, T.~Koenig, L.~Borner, and A.~Aurum, ``{An Analysis of Empirical
  Requirements Engineering Survey Data},'' in \emph{Engineering and Managing
  Software Requirements}.\hskip 1em plus 0.5em minus 0.4em\relax Springer,
  2005.

\bibitem{Glass.2002}
R.~L. Glass, I.~Vessey, and V.~Ramesh, ``{Research in Software Engineering: An
  Analysis of the Literature},'' \emph{Information and Software Technology},
  vol.~44, no.~8, 2002.

\bibitem{Jeffery.2002}
R.~Jeffery and L.~Scott, ``{Has Twenty-Five years of Empirical Software
  Engineering Made a Difference?}'' in \emph{Ninth Asia-Pacific Software
  Engineering Conference}.\hskip 1em plus 0.5em minus 0.4em\relax IEEE, 2002.

\bibitem{Segal.2005}
J.~Segal, A.~Grinyer, and H.~Sharp, ``{The Type of Evidence Produced by
  Empirical Software Engineers},'' in \emph{Workshop on Realising
  Evidence-Based Software Engineering}, 2005.

\bibitem{Zannier.2006}
C.~Zannier, G.~Melnik, and F.~Maurer, ``{On the Success of Empirical Studies in
  the International Conference on Software Engineering},'' in
  \emph{International Conference on Software Engineering}.\hskip 1em plus 0.5em
  minus 0.4em\relax Association for Computing Machinery, 2006.

\bibitem{Hoefer.2007}
A.~H{\"o}fer and W.~F. Tichy, ``{Status of Empirical Research in Software
  Engineering},'' in \emph{Empirical Software Engineering Issues. Critical
  Assessment and Future Directions}.\hskip 1em plus 0.5em minus 0.4em\relax
  Springer, 2007.

\bibitem{Borges.2015}
A.~Borges, W.~Ferreira, E.~Barreiros, A.~Almeida, L.~Fonseca \emph{et~al.},
  ``{Support Mechanisms to Conduct Empirical Studies in Software Engineering: A
  Systematic Mapping Study},'' in \emph{19th International Conference on
  Evaluation and Assessment in Software Engineering}, 2015.

\bibitem{Zhang.2018}
L.~Zhang, J.-H. Tian, J.~Jiang, Y.-J. Liu, M.-Y. Pu, and T.~Yue, ``{Empirical
  Research in Software Engineering - A Literature Survey},'' \emph{Journal of
  Computer Science and Technology}, vol.~33, no.~5, 2018.

\bibitem{Molleri.2019}
J.~S. Moll{\'e}ri, K.~Petersen, and E.~Mendes, ``{CERSE - Catalog for Empirical
  Research in Software Engineering: A Systematic Mapping Study},''
  \emph{Information and Software Technology}, vol. 105, 2019.

\bibitem{Guevara-Vega.2021}
C.~Guevara-Vega, B.~Bern{\'a}rdez, A.~Dur{\'a}n, A.~Quina-Mera, M.~Cruz, and
  A.~Ruiz-Cort{\'e}s, ``{Empirical Strategies in Software Engineering Research:
  A Literature Survey},'' in \emph{International Conference on Information
  Systems and Software Technologies}.\hskip 1em plus 0.5em minus 0.4em\relax
  IEEE, 2021.

\bibitem{Karras.2023}
\BIBentryALTinterwordspacing
O.~Karras, F.~Wernlein, J.~A.-C. Klünder, and S.~Auer,
  ``\BIBforeignlanguage{en}{{A Comparison of Scientific Publications on the
  State of Empirical Research in Requirements Engineering and Software
  Engineering}},'' 2023. [Online]. Available:
  \url{https://orkg.org/comparison/R255464/}
\BIBentrySTDinterwordspacing

\bibitem{Oelen.2021a}
A.~Oelen, M.~Stocker, and S.~Auer, ``{SmartReviews: Towards Human-and
  Machine-Actionable Representation of Review Articles},'' in
  \emph{International Conference on Asian Digital Libraries}.\hskip 1em plus
  0.5em minus 0.4em\relax Springer, 2021.

\bibitem{Oelen.2021b}
------, ``{SmartReviews: Towards Human-and Machine-Actionable Reviews},'' in
  \emph{International Conference on Theory and Practice of Digital
  Libraries}.\hskip 1em plus 0.5em minus 0.4em\relax Springer, 2021.

\bibitem{Bano.2014}
M.~Bano, D.~Zowghi, and N.~Ikram, ``{Systematic Reviews in Requirements
  Engineering: A Tertiary Study},'' in \emph{IEEE 4th International Workshop on
  Empirical Requirements Engineering}.\hskip 1em plus 0.5em minus 0.4em\relax
  IEEE, 2014.

\bibitem{dos_Santos.2021}
V.~{dos Santos}, A.~Y. Iwazaki, K.~R. Felizardo, {\'E}.~F. de~Souza, and E.~Y.
  Nakagawa, ``{Towards Sustainability of Systematic Literature Reviews},'' in
  \emph{15th ACM/IEEE International Symposium on Empirical Software Engineering
  and Measurement}, 2021.

\bibitem{Mendes.2020}
E.~Mendes, C.~Wohlin, K.~Felizardo, and M.~Kalinowski, ``{When to Update
  Systematic Literature Reviews in Software Engineering},'' \emph{Journal of
  Systems and Software}, vol. 167, 2020.

\bibitem{Wohlin.2020}
C.~Wohlin, E.~Mendes, K.~R. Felizardo \emph{et~al.}, ``{Guidelines for the
  Search Strategy to Update Systematic Literature Reviews in Software
  Engineering},'' \emph{Information and Software Technology}, vol. 127, 2020.

\bibitem{Felizardo.2020}
K.~R. Felizardo, {\'E}.~F. de~Souza, T.~Malacrida, B.~M. Napole{\~a}o,
  F.~Petrillo \emph{et~al.}, ``{Knowledge Management for Promoting Update of
  Systematic Literature Reviews: An Experience Report},'' in \emph{46th
  Euromicro Conference on Software Engineering and Advanced Applications},
  2020.

\bibitem{Mendez.2020}
D.~Mendez, D.~Graziotin, S.~Wagner, and H.~Seibold, ``{Open Science in Software
  Engineering},'' in \emph{Contemporary Empirical Methods in Software
  Engineering}.\hskip 1em plus 0.5em minus 0.4em\relax Springer, 2020.

\bibitem{Wilkinson.2016}
M.~D. Wilkinson, M.~Dumontier, I.~J. Aalbersberg, G.~Appleton, M.~Axton
  \emph{et~al.}, ``{The FAIR Guiding Principles for Scientific Data Management
  and Stewardship},'' \emph{Scientific Data}, vol.~3, no.~1, 2016.

\bibitem{Stocker.2023}
M.~Stocker, A.~Oelen, M.~Y. Jaradeh, M.~Haris, O.~A. Oghli \emph{et~al.},
  ``{FAIR Scientific Information with the Open Research Knowledge Graph},''
  \emph{FAIR Connect}, vol.~1, no.~1, 2023.

\bibitem{Auer.2020}
S.~Auer, A.~Oelen, M.~Haris, M.~Stocker, J.~D’Souza, K.~E. Farfar, L.~Vogt,
  M.~Prinz, V.~Wiens, and M.~Y. Jaradeh, ``{Improving Access to Scientific
  Literature with Knowledge Graphs},'' \emph{Bibliothek Forschung und Praxis},
  vol.~44, no.~3, 2020.

\bibitem{Auer.2018}
S.~Auer, V.~Kovtun, M.~Prinz, A.~Kasprzik, M.~Stocker, and M.~E. Vidal,
  ``{Towards a Knowledge Graph for Science},'' in \emph{8th International
  Conference on Web Intelligence, Mining and Semantics}, 2018.

\bibitem{Dessi.2022}
D.~Dess{\'\i}, F.~Osborne, D.~Reforgiato~Recupero, D.~Buscaldi, and E.~Motta,
  ``{CS-KG: A Large-Scale Knowledge Graph of Research Entities and Claims in
  Computer Science},'' in \emph{International Semantic Web Conference}.\hskip
  1em plus 0.5em minus 0.4em\relax Springer, 2022.

\bibitem{Hussein.2022}
H.~Hussein, A.~Oelen, O.~Karras, and S.~Auer, ``{KGMM - A Maturity Model for
  Scholarly Knowledge Graphs Based on Intertwined Human-Machine
  Collaboration},'' in \emph{From Born-Physical to Born-Virtual: Augmenting
  Intelligence in Digital Libraries: 24th International Conference on Asian
  Digital Libraries}.\hskip 1em plus 0.5em minus 0.4em\relax Springer, 2022.

\bibitem{Auer.2007}
S.~Auer, C.~Bizer, G.~Kobilarov, J.~Lehmann \emph{et~al.}, ``{DBpedia: A
  Nucleus for a Web of Open Data},'' in \emph{The Semantic Web}.\hskip 1em plus
  0.5em minus 0.4em\relax Springer, 2007.

\bibitem{Vrandevcic.2014}
D.~Vrande{\v{c}}i{\'c} and M.~Kr{\"o}tzsch, ``{Wikidata: A Free Collaborative
  Knowledge Base},'' \emph{Communications of the ACM}, vol.~57, no.~10, 2014.

\bibitem{Ammar.2018}
W.~Ammar, D.~Groeneveld, C.~Bhagavatula, I.~Beltagy, M.~Crawford \emph{et~al.},
  ``{Construction of the Literature Graph in Semantic Scholar},'' in
  \emph{Conference of the North {A}merican Chapter of the Association for
  Computational Linguistics: Human Language Technologies, Volume 3}.\hskip 1em
  plus 0.5em minus 0.4em\relax Association for Computational Linguistics, 2018.

\bibitem{Jaradeh.2019}
M.~Y. Jaradeh, A.~Oelen, K.~E. Farfar, M.~Prinz, J.~D'Souza, G.~Kismih{\'o}k,
  M.~Stocker, and S.~Auer, ``{Open Research Knowledge Graph: Next Generation
  Infrastructure for Semantic Scholarly Knowledge},'' in \emph{10th
  International Conference on Knowledge Capture}, 2019.

\bibitem{Jeschke.2020}
\BIBentryALTinterwordspacing
J.~Jeschke, M.~Enders, M.~Bagni, D.~Aumann, P.~Jeschke, M.~Zimmermann, and
  T.~Heger. (2020) {Hi-Knowledge.org}. [Online]. Available:
  \url{https://hi-knowledge.org/}
\BIBentrySTDinterwordspacing

\bibitem{Paperswithcode.2022}
\BIBentryALTinterwordspacing
(2020) {Papers With Code}. [Online]. Available:
  \url{https://paperswithcode.com/about}
\BIBentrySTDinterwordspacing

\bibitem{Penev.2019}
L.~Penev, M.~Dimitrova, V.~Senderov, G.~Zhelezov, T.~Georgiev \emph{et~al.},
  ``{OpenBiodiv: A Knowledge Graph for Literature-Extracted Linked Open Data in
  Biodiversity Science},'' \emph{Publications}, vol.~7, no.~2, 2019.

\bibitem{Gkatzelis.2021}
G.~I. Gkatzelis, J.~B. Gilman, S.~S. Brown, H.~Eskes, A.~R. Gomes, A.~C. Lange,
  B.~C. McDonald, J.~Peischl, A.~Petzold, C.~R. Thompson, and
  A.~Kiendler-Scharr, ``{The Global Impacts of COVID-19 Lockdowns on Urban Air
  Pollution: A Critical Review and Recommendations},'' \emph{Elementa: Science
  of the Anthropocene}, vol.~9, no.~1, 2021.

\bibitem{Spadaro.2022}
G.~Spadaro, I.~Tiddi, S.~Columbus, S.~Jin, A.~ten Teije \emph{et~al.}, ``{The
  Cooperation Databank: Machine-Readable Science Accelerates Research
  Synthesis},'' \emph{Perspectives on Psychological Science}, 2022.

\bibitem{Stocker.2022}
M.~Stocker, T.~Heger, A.~Schweidtmann, H.~{\'C}wiek-Kupczy{\'n}ska, L.~Penev,
  M.~Dojchinovski, E.~Willighagen, M.-E. Vidal, H.~Turki, D.~Balliet
  \emph{et~al.}, ``{SKG4EOSC - Scholarly Knowledge Graphs for EOSC:
  Establishing a Backbone of Knowledge Graphs for FAIR Scholarly Information in
  EOSC},'' \emph{Research Ideas and Outcomes}, vol.~8, 2022.

\bibitem{Frattini.2022}
J.~Frattini, L.~Montgomery, J.~Fischbach, M.~Unterkalmsteiner, D.~Mendez, and
  D.~Fucci, ``{A Live Extensible Ontology of Quality Factors for Textual
  Requirements},'' in \emph{IEEE 30th International Requirements Engineering
  Conference}, 2022.

\bibitem{Karras.2021}
O.~Karras, E.~C. Groen, J.~A. Khan, and S.~Auer, ``{Researcher or Crowd Member?
  Why not both! The Open Research Knowledge Graph for Applying and
  Communicating CrowdRE Research},'' in \emph{29th International Requirements
  Engineering Conference Workshops}.\hskip 1em plus 0.5em minus 0.4em\relax
  IEEE, 2021.

\bibitem{Karras.2021a}
\BIBentryALTinterwordspacing
O.~Karras and E.~C. Groen, ``{Overview of Approaches that Classify User
  Feedback as Feature Request},'' Open Research Knowledge Graph, 2021.
  [Online]. Available: \url{https://doi.org/10.48366/R112387}
\BIBentrySTDinterwordspacing

\bibitem{Karras2021b}
\BIBentryALTinterwordspacing
O.~Karras and J.~A. Khan, ``{Overview of Crowd Intelligence in Requirements
  Engineering},'' Open Research Knowledge Graph, 2021. [Online]. Available:
  \url{https://doi.org/10.48366/R114155}
\BIBentrySTDinterwordspacing

\bibitem{Santos.2019}
R.~Santos, E.~C. Groen, and K.~Villela, ``{An Overview of User Feedback
  Classification Approaches},'' in \emph{International Working Conference on
  Requirements Engineering: Foundation for Software Quality, Workshops}, 2019.

\bibitem{Khan.2019}
J.~A. Khan, L.~Liu, L.~Wen, and R.~Ali, ``{Crowd Intelligence in Requirements
  Engineering: Current Status and Future Directions},'' in \emph{International
  Working Conference on Requirements Engineering: Foundation for Software
  Quality}.\hskip 1em plus 0.5em minus 0.4em\relax Springer, 2019.

\bibitem{Abualhaija.2022}
S.~Abualhaija, C.~Arora, A.~Sleimi, and L.~C. Briand, ``{Automated Question
  Answering for Improved Understanding of Compliance Requirements: A
  Multi-Document Study},'' in \emph{IEEE 30th International Requirements
  Engineering Conference}, 2022.

\bibitem{Gruninger.1995}
M.~Gr\"uninger and M.~S. Fox, ``{The Role of Competency Questions in Enterprise
  Engineering},'' in \emph{{Benchmarking -- Theory and Practice}}.\hskip 1em
  plus 0.5em minus 0.4em\relax Springer, 1995.

\bibitem{Hogan.2021}
\BIBentryALTinterwordspacing
A.~Hogan, E.~Blomqvist, M.~Cochez, C.~d'Amato, G.~de~Melo, C.~Guti\'errez,
  S.~Kirrane, J.~E. Labra~Gayo, R.~Navigli, S.~Neumaier, A.-C. Ngonga~Ngomo,
  A.~Polleres, S.~M. Rashid, A.~Rula, L.~Schmelzeisen, J.~F. Sequeda, S.~Staab,
  and A.~Zimmermann, \emph{{K}nowledge {G}raphs}, ser. Synthesis Lectures on
  Data, Semantics, and Knowledge.\hskip 1em plus 0.5em minus 0.4em\relax Morgan
  \& Claypool, 2021. [Online]. Available: \url{https://kgbook.org/}
\BIBentrySTDinterwordspacing

\bibitem{Sjoberg.2007}
D.~I. Sjoberg, T.~Dyba, and M.~Jorgensen, ``{The Future of Empirical Methods in
  Software Engineering Research},'' in \emph{Future of Software
  Engineering}.\hskip 1em plus 0.5em minus 0.4em\relax IEEE, 2007.

\bibitem{Wiens.2020}
V.~Wiens, M.~Stocker, and S.~Auer, ``{Towards Customizable Chart Visualizations
  of Tabular Data Using Knowledge Graphs},'' in \emph{Digital Libraries at
  Times of Massive Societal Transition}.\hskip 1em plus 0.5em minus 0.4em\relax
  Springer, 2020.

\bibitem{Jaradeh.2020}
M.~Y. Jaradeh, M.~Stocker, and S.~Auer, ``{Question Answering on Scholarly
  Knowledge Graphs},'' in \emph{International Conference on Theory and Practice
  of Digital Libraries}.\hskip 1em plus 0.5em minus 0.4em\relax Springer, 2020.

\bibitem{Auer.2023}
S.~Auer, D.~A.~C. Barone, C.~Bartz, E.~G. Cortes, M.~Y. Jaradeh, O.~Karras,
  M.~Koubarakis, D.~Mouromtsev, D.~Pliukhin, D.~Radyush, I.~Shilin, M.~Stocker,
  and E.~Tsalapati, ``{The SciQA Scientific Question Answering Benchmark for
  Scholarly Knowledge},'' \emph{Nature Scientific Reports}, vol.~13, no. 7240,
  2023.

\bibitem{Faerber.2019}
M.~F{\"a}rber, ``{The Microsoft Academic Knowledge Graph: A Linked Data Source
  with 8 Billion Triples of Scholarly Data},'' in \emph{International Semantic
  Web Conference}.\hskip 1em plus 0.5em minus 0.4em\relax Springer, 2019.

\bibitem{Priem.2022}
J.~Priem, H.~Piwowar, and R.~Orr, ``{OpenAlex: A Fully-Open Index of Scholarly
  Works, Authors, Venues, Institutions, and Concepts},'' \emph{arXiv preprint
  arXiv:2205.01833}, 2022.

\bibitem{Hammond.2017}
T.~Hammond, M.~Pasin, and E.~Theodoridis, ``{Data Integration and
  Disintegration: Managing Springer Nature SciGraph with SHACL and OWL},'' in
  \emph{International Semantic Web Conference}, 2017.

\bibitem{Manghi.2019}
P.~Manghi, A.~Bardi, C.~Atzori, M.~Baglioni, N.~Manola, J.~Schirrwagen,
  P.~Principe, M.~Artini, A.~Becker, M.~De~Bonis \emph{et~al.}, ``{The OpenAIRE
  Research Graph Data Model},'' \emph{Zenodo}, 2019.

\bibitem{Schirrwagen.2013}
J.~Schirrwagen, P.~Manghi, N.~Manola, L.~Bolikowski, N.~Rettberg, and
  B.~Schmidt, ``{Data Curation in the OpenAIRE Scholarly Communication
  Infrastructure},'' \emph{Information Standards Quarterly}, vol.~25, no.~3,
  2013.

\bibitem{Aryani.2017}
A.~Aryani and J.~Wang, ``{Research Graph: Building a Distributed Graph of
  Scholarly Works Using Research Data Switchboard},'' 2017.

\bibitem{Burton.2017}
A.~Burton, H.~Koers, P.~Manghi, M.~Stocker, M.~Fenner, A.~Aryani, S.~La~Bruzzo,
  M.~Diepenbroek, and U.~Schindler, ``{The Scholix Framework for
  Interoperability in Data-Literature Information Exchange},'' \emph{D-Lib
  Magazine}, vol.~23, no. 1/2, 2017.

\bibitem{Brack.2022}
A.~Brack, A.~Hoppe, M.~Stocker, S.~Auer, and R.~Ewerth, ``{Analysing the
  Requirements for an Open Research Knowledge Graph: Use Cases, Quality
  Requirements, and Construction Strategies},'' \emph{International Journal on
  Digital Libraries}, vol.~23, no.~1, 2022.

\bibitem{Domingo-Fernandez.2020}
D.~Domingo-Fernández, S.~Baksi, B.~Schultz, Y.~Gadiya, R.~Karki, T.~Raschka,
  C.~Ebeling, M.~Hofmann-Apitius, and A.~T. Kodamullil, ``{COVID-19 Knowledge
  Graph: A Computable, Multi-Modal, Cause-and-Effect Knowledge Model of
  COVID-19 Pathophysiology},'' \emph{Bioinformatics}, vol.~37, no.~9, 12 2020.

\bibitem{Covid-19_air_quality.2021}
\BIBentryALTinterwordspacing
(2021) {COVID-19 Air Quality Data Collection}. [Online]. Available:
  \url{https://covid-aqs.fz-juelich.de}
\BIBentrySTDinterwordspacing

\bibitem{Schindler.2020}
D.~Schindler, B.~Zapilko, and F.~Kr{\"u}ger, ``{Investigating Software Usage in
  the Social Sciences: A Knowledge Graph Approach},'' in \emph{European
  Semantic Web Conference}.\hskip 1em plus 0.5em minus 0.4em\relax Springer,
  2020.

\bibitem{Schindler.2021}
D.~Schindler, F.~Bensmann, S.~Dietze, and F.~Kr{\"u}ger, ``{SoMeSci - A 5 Star
  Open Data Gold Standard Knowledge Graph of Software Mentions in Scientific
  Articles},'' in \emph{30th ACM International Conference on Information \&
  Knowledge Management}, 2021.

\bibitem{Schindler.2022}
------, ``{The Role of Software in Science: A Knowledge Graph-Based Analysis of
  Software Mentions in PubMed Central},'' \emph{PeerJ Computer Science},
  vol.~8, 2022.

\bibitem{Sjoberg.2005}
D.~I. Sj{\o}berg, J.~E. Hannay, O.~Hansen, V.~B. Kampenes, A.~Karahasanovic,
  N.-K. Liborg, and A.~C. Rekdal, ``{A Survey of Controlled Experiments in
  Software Engineering},'' \emph{IEEE Transactions on Software Engineering},
  vol.~31, no.~9, 2005.

\bibitem{Bezerra.2015}
R.~M. Bezerra, F.~Q. da~Silva, A.~M. Santana, C.~V. Magalhaes, and R.~E.
  Santos, ``{Replication of Empirical Studies in Software Engineering: An
  Update of a Systematic Mapping Study},'' in \emph{ACM/IEEE International
  Symposium on Empirical Software Engineering and Measurement}.\hskip 1em plus
  0.5em minus 0.4em\relax IEEE, 2015.

\bibitem{Dyba.2008}
T.~Dyb{\aa} and T.~Dings{\o}yr, ``{Empirical Studies of Agile Software
  Development: A Systematic Review},'' \emph{Information and Software
  Technology}, vol.~50, no. 9-10, 2008.

\bibitem{Zhang.2016}
T.~Zhang, H.~Jiang, X.~Luo, and A.~T. Chan, ``{A Literature Review of Research
  in Bug Resolution: Tasks, Challenges and Future Directions},'' \emph{The
  Computer Journal}, vol.~59, no.~5, 2016.

\bibitem{Guevara.2021a}
\BIBentryALTinterwordspacing
C.~Guevara-Vega, B.~Bernárdez, A.~Durán, A.~Quiña-Mera, M.~Cruz, and
  A.~Ruiz-Cortés, ``{80 Initial Data-set Studies SMS Strategy CG},'' 2021.
  [Online]. Available: \url{https://doi.org/10.5281/zenodo.4456034}
\BIBentrySTDinterwordspacing

\bibitem{Guevara.2021b}
\BIBentryALTinterwordspacing
------, ``{20 Primary Studies SMS Strategy CG},'' 2021. [Online]. Available:
  \url{https://doi.org/10.5281/zenodo.4455951}
\BIBentrySTDinterwordspacing

\bibitem{Ralph.2021}
P.~Ralph, N.~b. Ali, S.~Baltes, D.~Bianculli, J.~Diaz, Y.~Dittrich, N.~Ernst,
  M.~Felderer, R.~Feldt, A.~Filieri \emph{et~al.}, ``{Empirical Standards for
  software engineering research},'' \emph{arXiv preprint arXiv:2010.03525},
  2021.

\bibitem{Basili.1994}
V.~R. Basili, C.~Caldiera, and H.~D. Rombach, ``{Goal Question Metric
  Paradigm},'' \emph{Encyclopedia of Software Engineering}, vol.~1, 1994.

\bibitem{Runeson.2020}
P.~Runeson, E.~Engstr{\"o}m, and M.-A. Storey, ``{The Design Science Paradigm
  as a Frame for Empirical Software Engineering},'' \emph{{Contemporary
  Empirical Methods in Software Engineering}}, 2020.

\bibitem{Karras.2023a}
\BIBentryALTinterwordspacing
O.~Karras, ``{Analysis of the State and Evolution of Empirical Research in
  Requirements Engineering},'' 2023. [Online]. Available:
  \url{https://doi.org/10.5281/zenodo.8083529}
\BIBentrySTDinterwordspacing

\bibitem{Kabongo.2021}
S.~Kabongo, J.~D’Souza, and S.~Auer, ``{Automated Mining of Leaderboards for
  Empirical AI Research},'' in \emph{International Conference on Asian Digital
  Libraries}.\hskip 1em plus 0.5em minus 0.4em\relax Springer, 2021.

\bibitem{Knublauch.2017}
\BIBentryALTinterwordspacing
H.~Knublauch and D.~Kontokostas, ``{Shapes Constraint Language (SHACL)},'' W3C,
  {W3C Recommendation}, 2017. [Online]. Available:
  \url{https://www.w3.org/TR/2017/REC-shacl-20170720/}
\BIBentrySTDinterwordspacing

\bibitem{Seaborne.2013}
\BIBentryALTinterwordspacing
A.~Seaborne and S.~Harris, ``{SPARQL 1.1 Query Language},'' W3C, {W3C
  Recommendation}, 2013. [Online]. Available:
  \url{https://www.w3.org/TR/2013/REC-sparql11-query-20130321/}
\BIBentrySTDinterwordspacing

\bibitem{Wohlin.2021}
C.~Wohlin, ``{Case Study Research in Software Engineering -- It is a Case, and
  it is a Study, but is it a Case Study?}'' \emph{Information and Software
  Technology}, vol. 133, 2021.

\bibitem{Wilcoxon.1945}
F.~Wilcoxon, ``{Individual Comparisons by Ranking Methods},'' \emph{Biometrics
  Bulletin}, vol.~1, no.~6, 1945.

\bibitem{Mann.1947}
H.~B. Mann and D.~R. Whitney, ``{On a Test of Whether One of Two Random
  Variables is Stochastically Larger than the Other},'' \emph{The Annals of
  Mathematical Statistics}, vol.~18, no.~1, 1947.

\bibitem{Ampatzoglou.2020}
A.~Ampatzoglou, S.~Bibi, P.~Avgeriou, and A.~Chatzigeorgiou, \emph{{Guidelines
  for Managing Threats to Validity of Secondary Studies in Software
  Engineering}}.\hskip 1em plus 0.5em minus 0.4em\relax Springer, 2020.

\bibitem{Glinz.2015}
M.~Glinz and S.~A. Fricker, ``{On Shared Understanding in Software Engineering:
  An Essay},'' \emph{Computer Science - Research and Development}, vol.~30,
  2015.

\bibitem{Karras.2020}
O.~Karras, K.~Schneider, and S.~A. Fricker, ``{Representing Software Project
  Vision by Means of Video: A Quality Model for Vision Videos},'' \emph{Journal
  of Systems and Software}, vol. 162, 2020.

\bibitem{Jinha.2010}
A.~E. Jinha, ``{Article 50 Million: An Estimate of the Number of Scholarly
  Articles in Existence},'' \emph{Learned Publishing}, vol.~23, no.~3, 2010.

\bibitem{White.2019}
K.~White, ``{Publications Output: US Trends and International Comparisons.
  Science \& Engineering Indicators 2020. NSB-2020-6.}'' \emph{National Science
  Foundation}, 2019.

\bibitem{Martin.2022}
L.~Martin and A.~Henrich, ``{RDFtex: Knowledge Exchange Between LaTeX-Based
  Research Publications and Scientific Knowledge Graphs},'' in \emph{Linking
  Theory and Practice of Digital Libraries}.\hskip 1em plus 0.5em minus
  0.4em\relax Springer, 2022.

\bibitem{Bless.2023}
C.~Bless, I.~Baimuratov, and O.~Karras, ``{SciKGTeX - A LaTeX Package to
  Semantically Annotate Contributions in Scientific Publications},'' in
  \emph{23nd ACM/IEEE Joint Conference on Digital Libraries}, 2023.

\end{thebibliography}

\end{document}